\documentclass[aps,pra,twocolumn,showpacs,superscriptaddress,floatfix,notitlepage,showkeys]{revtex4-1}
\usepackage{amssymb,amsmath,amsfonts}
\usepackage{epsfig,graphicx}
\usepackage{afterpage}
\usepackage{hyperref}
\hypersetup{colorlinks,citecolor=PineGreen,filecolor=black,linkcolor=black,urlcolor=black}

\newcommand\adag{a^\dagger}
\newcommand\adaga{a^\dagger a}
\newcommand\ket[1]{\left|\textstyle{#1}\right\rangle}
\newcommand\bra[1]{\left\langle\textstyle{#1}\right|}

\usepackage{color}
\definecolor{blueRP}{rgb}{0.0, 0.4, 1.}
\definecolor{blueCh}{rgb}{0.2, 0.4, 0.7}

\usepackage[dvipsnames]{xcolor}

\begin{document}

\title{Connecting nth order generalised quantum Rabi models: Emergence of nonlinear spin-boson coupling via spin rotations}
\author{Jorge Casanova}\email{jcasanovamar@gmail.com}
\affiliation{Institut f\"{u}r Theoretische Physik and IQST, Albert-Einstein Allee 11, Universit\"{a}t Ulm, D-89069 Ulm, Germany}
\author{Ricardo Puebla}\email{rpueblaantunes@gmail.com}
\affiliation{Institut f\"{u}r Theoretische Physik and IQST, Albert-Einstein Allee 11, Universit\"{a}t Ulm, D-89069 Ulm, Germany}
\affiliation{Centre for Theoretical Atomic, Molecular, and Optical Physics, School of Mathematics and Physics, Queen's University, Belfast BT7 1NN, United Kingdom}
\author{Hector Moya-Cessa}
\affiliation{Instituto Nacional de Astrof\'isica, \'Optica y Electr\'onica, Calle Luis Enrique Erro 1, Santa Mar\'ia Tonantzintla, Puebla, 72840 Mexico}\affiliation{Institut f\"ur Quantenphysik and IQST, Universit\"at Ulm, Albert-Einstein-Allee 11, D-89081 Ulm, Germany}
\author{Martin B. Plenio}
\affiliation{Institut f\"{u}r Theoretische Physik and IQST, Albert-Einstein Allee 11, Universit\"{a}t Ulm, D-89069 Ulm, Germany}

\begin{abstract}
 We establish an approximate equivalence between a generalised quantum Rabi model and its nth order counterparts where spin-boson interactions are nonlinear as they comprise a simultaneous exchange of $n$ bosonic excitations. Although there exists no unitary transformation between these models, we demonstrate their equivalence to a good approximation in a wide range of parameters. This shows that nonlinear spin-boson couplings, i.e. nth order quantum Rabi models, are accessible to quantum systems with only linear coupling between boson and spin modes by simply adding spin rotations and after an appropriate transformation. Furthermore, our result prompts novel approximate analytical solutions to the dynamics of the quantum Rabi model in the ultrastrong coupling regime improving previous approaches. 
\end{abstract}

\maketitle

\section*{Introduction}

The quantum Rabi model (QRM) lies not only at the heart of our understanding of light-matter interaction~\cite{Scully}, but is also of importance in diverse fields of research~\cite{Braak:16}. The Rabi model was primarily
proposed to describe a nuclear spin interacting with classical
radiation~\cite{Rabi:36,Rabi:37}, whose quantised version
only appeared two decades later~\cite{Jaynes:63}. This contemplates a scenario which is of great generality  as it encompasses two of the
most basic, yet essential, ingredients in quantum physics, namely, a
two-level system and a bosonic mode. Indeed, this model emerges in disparate settings, ranging from
ion traps~\cite{Leibfried:03,Haffner:08} to circuit or cavity
QED~\cite{Haroche,Devoret:13}, quantum optomechanical systems~\cite{Aspelmeyer:14}, color-centers in membranes~\cite{Abdi:17}, and cold atoms~\cite{Schneeweiss:17}.

Even though the QRM has been exhaustively investigated in the
last decades, a number of recent findings has brought it again into the
research spotlight. Among them we can mention its integrability~\cite{Braak:11}, the existence of a distinctive behaviour in the deep strong coupling regime~\cite{Casanova:10}, or the emergence of a
quantum phase transition~\cite{Hwang:15,Puebla:16,Puebla:17,Puebla:16njp}.
Closely related to the QRM, we find the nth order QRM (nQRM)
which differs from the QRM in that the nQRM comprises $n$-boson
exchange interaction terms because of the presence of a nonlinear spin-boson coupling. This generalisation of the QRM
has recently attracted attention, mainly in its second-order form
(2QRM) as it shows striking phenomena such as spectral
collapse~\cite{Felicetti:15,Duan:16,Puebla:17pra}, due to its relevance in preparing non-classical states of light in quantum optics~\cite{Brune:87,Toor:92} and regarding its solvability~\cite{Travenec:12,Chen:12,Cui:17}. These studies have also been extended to a mixed QRM comprising both one- and two-boson interaction terms, which appears in the context of circuit QED~\cite{Bertet:05,Bertet:05a,Felicetti:18}. Furthermore, solutions to this mixed QRM have recently been found~\cite{Duan:18}, and it has also been reported that this model displays quantum phase transitions~\cite{Ying:18}. Due to these compelling physical properties, the coherent control of nth order quantum Rabi models could open new avenues to develop different fields as quantum computing or quantum simulations. In addition, because of their different spectra, it is worth noting that there is no unitary map between the QRM and the nQRM with $n>1$.

In this article, we demonstrate the existence of a connection, i.e. an approximate equivalence, among a family of Hamiltonians comprising nth order boson interaction terms, where the standard QRM or the  2QRM appear as special cases. As a proof of concept, we show how the dynamics of a 2QRM and a 3QRM can be captured without having access to the required nonlinear two- and three-photon interactions, and after an appropriate transformation of a linear QRM that includes spin driving terms, i.e. spin rotations. The latter is dubbed here as generalised QRM (gQRM). In this manner we can argue that, a quantum system that contains a linear spin-boson coupling but lacks of nonlinear interactions suffices for the simulation of models where nonlinear terms are crucial. Our method works as follows: The dynamics of a state $\ket{\psi}$ evolving under a nQRM (the targeted dynamics) can be successfully retrieved from a gQRM (the starting point of our method) by i) evolving a transformed initial state $T\ket{\psi}$ under gQRM  during a time $t$ and ii) measuring customary spin and boson observables of the gQRM. We will demonstrate that the latter corresponds to expectation values of observables of the state $\ket{\psi(t)}$ evolved under the nonlinear nQRM (see Fig.~\ref{fig0} for a scheme of the method). Indeed, as creating $n$-boson interactions is considered challenging in many quantum platforms, our method opens new avenues for their inspection.
It is worth stressing that this reported method fundamentally differs from previous works where resonant multi-boson effective Hamiltonians were obtained, either via amplitude modulation as used in circuit QED~\cite{Strand:13,Allman:14,Lu:17}, or via adiabatic passage~\cite{Ma:15,Garziano:15}. In these works effective multi-boson exchange terms do not comprise nonlinear spin-boson couplings and hold only in a very limited parameter regime and/or for particular states. Certainly, in this article we report an approximate equivalence among nQRMs which holds in a large range of parameters and  grants a large tunability to explore their physics, as well as it unveils a fundamental relation between these  models.
Moreover, we present a potentially scalable platform~\cite{Lekitsch:17}, a microwave-driven trapped ion setting~\cite{Mintert:01,Timoney:11,Weidt:16,Piltz:16}, in which nQRMs are unattainable without resorting to our approximate equivalence, which highlights the applicability of our method.  Finally, we use our theory to analyse the standard QRM and find that our method provides, in addition, approximate analytical solutions that surpass in accuracy previous approaches in the ultrastrong coupling regime~\cite{Forn:10,Beaudoin:11,Rossatto:16, Feranchuk:96,Irish:07,Gan:10}.

\section*{Results}

\subsection*{Description of the approximate equivalence}

\begin{figure}
\centering
\includegraphics[width=1\linewidth,angle=-00]{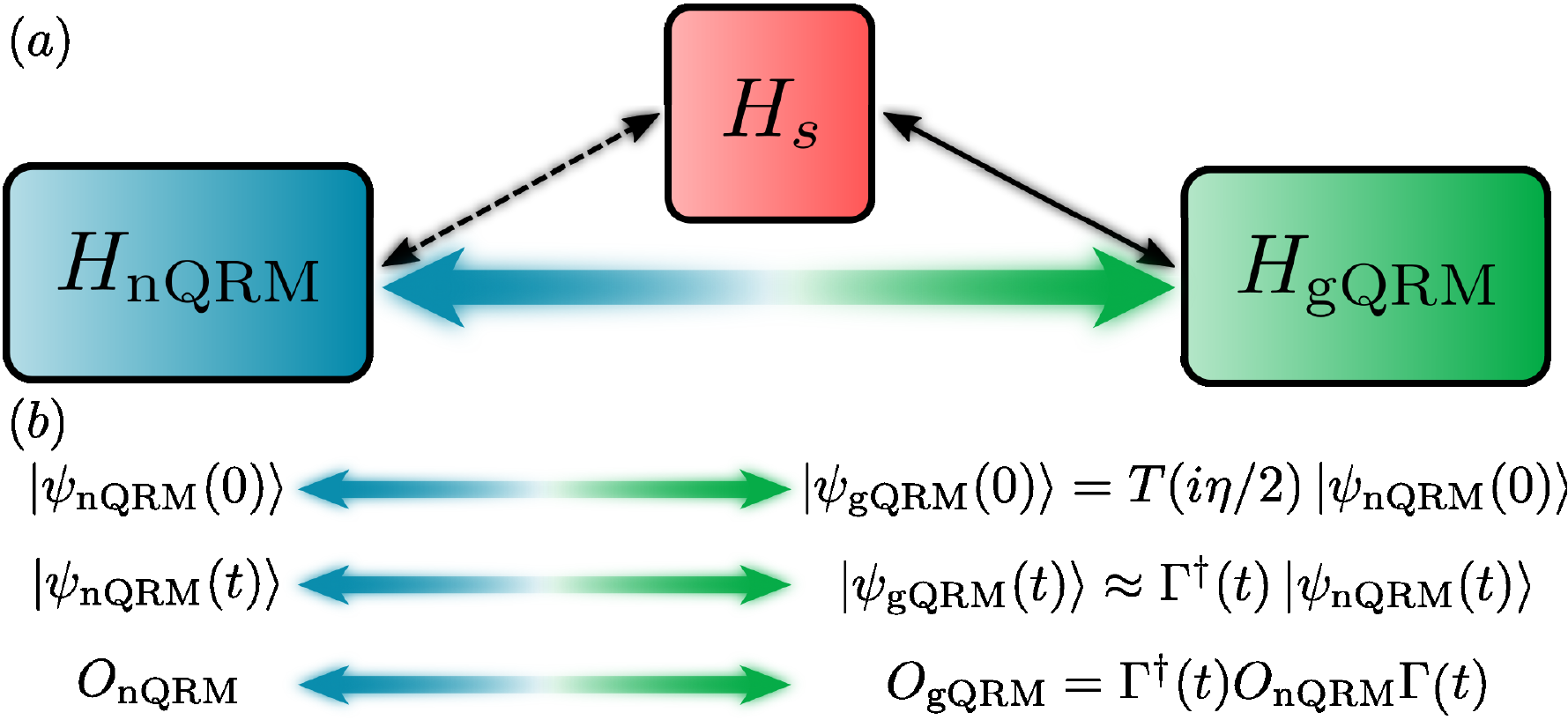}
\caption{\small{Scheme of the approximate equivalence. (a) Diagram of the approximate equivalence among $H_s$, $H_{\rm nQRM}$ and $H_{\rm gQRM}$. The transformation between $H_s$ and $H_{\rm gQRM}$ is exact (solid arrow), while between $H_s$ and $H_{\rm nQRM}$ is approximated (dashed arrow). Hence, we establish an approximate map between nQRM and gQRM (blurred arrow). (b) The latter is accomplished by the transformations between initial states ($|\psi_{\rm nQRM}(0)\rangle$, $|\psi_{\rm gQRM}(0)\rangle$), evolved states ($|\psi_{\rm nQRM}(t)\rangle$, $|\psi_{\rm gQRM}(t)\rangle$), and observables ($O_{\rm nQRM}, O_{\rm gQRM} $).}}
\label{fig0}
\end{figure}

We begin with the following general Hamiltonian (later we will demonstrate its connection with the gQRM that only contains linear spin-boson interactions and represents the starting point of our approximate equivalence)
\begin{equation}
\label{eq:Hs}
H_{s}=\nu\adaga+\frac{\omega}{2}\sigma_z +\frac{\Omega}{2}\sum_j\left[ \sigma^+ e^{i\eta(a+\adag)}e^{-i\alpha_j}+\rm{H.c}\right],
\end{equation}
whose first two terms correspond to a bosonic mode of frequency $\nu$ and a two-level system with a frequency splitting $\omega$, described by the usual annihilation (creation) operator $a$ ($\adag$) and spin-$\frac{1}{2}$ Pauli matrices $\vec{\sigma}=\left(\sigma_x,\sigma_y,\sigma_z\right)$, respectively. Both subsystems interact through a set of coupling terms with amplitude $\Omega/2$ and parameter $\eta$, considered here equal  $\forall j$, and $\alpha_j$ being a time dependent phase. The Hamiltonian $H_s$ is central for our theory, as sketched in Fig.~\ref{fig0}, and establishes an approximate map between gQRM dynamics with those of the nQRM.  We perform a unitary transformation on $H_s$ to find $H_{T}=T(i\eta/2)H_{s}T^{\dagger}(i\eta/2)$, where $T(\beta)=1/\sqrt{2}\left[\mathcal{D}(\beta)\left(\ket{e}\bra{g}+\ket{g}\bra{g}\right)+\mathcal{D}^{\dagger}(\beta)\left(\ket{e}\bra{e}-\ket{g}\bra{e}\right) \right]$ with $\sigma_z=\ket{e}\bra{e}-\ket{g}\bra{g}$
and $\mathcal{D}(\beta)=e^{\beta \adag-\beta^{*}a}$ is the displacement operator. Note that this transformation has been used in the context of trapped ions  to derive the eigenstates of a system that comprises a laser interacting with a trapped ion, and for  fast implementations of the QRM~\cite{MoyaCessa:03,MoyaCessa:16}. Now, choosing time dependent phases, $\alpha_j=(\omega +\delta_j)t$, and moving to a rotating frame with respect to $H_{T,0}=-(\omega+\delta_1) \sigma_x/2$, the resulting Hamiltonian, $H_{\rm gQRM}$, reads  (for more details see Methods section)
\begin{widetext}
\begin{align}
H_{\rm gQRM}&\equiv \mathcal{U}_{T,0}^{\dagger}(t)(H_{T}-H_{T,0})\mathcal{U}_{T,0}(t)=\nu \adaga+\frac{\delta_1}{2}\sigma_x-\frac{\eta \nu }{2}p\sigma_x+\frac{\Omega}{2}\sum_{j}\bigg\{\cos[ (\delta_j-\delta_1)t]\sigma_z+\sin[ (\delta_j-\delta_1)t]\sigma_y \bigg\} \label{eq:gQRM}
\end{align}
\end{widetext}
with $p=i(\adag-a)$ and $\mathcal{U}_{T,0}=e^{-itH_{T,0}}$. The previous Hamiltonian is the one of the gQRM, where its last term can be viewed as a classical driving acting on the system, i.e. this is the term leading to spin rotations.  In particular, we note that  $H_{\rm gQRM}$ adopts the form of a standard QRM in the case of having $\delta_{j}=0 \ \forall j $.

On the other hand, the Hamiltonian $H_s$ in Eq.~(\ref{eq:Hs}) can be brought into the form of a $H_{\rm nQRM}$ by properly choosing $\alpha_j$ and in a suitable interaction picture. More specifically, by defining $H_s=H_{s,0}+H_{s,1}$ with $H_{s,0}=(\nu-\tilde{\nu})\adaga+(\omega-\tilde{\omega})\sigma_z/2$ and considering two interaction  terms (i.e. $j=1,2$) such that $\delta_{1,2}=\mp n\nu-\tilde{\omega}\pm n\tilde{\nu}$ (recall that $\alpha_j=(\omega +\delta_j) t$ and thus $\alpha_j$ and $\delta_j$ are related)  with $\tilde{\omega}>0$ and $\tilde{\nu}>0$, we find that $H_{s,1}^I = e^{itH_{s,0}} H_{s,1}e^{-itH_{s,0}}$ approximately leads to 
\begin{equation}\label{eq:nQRM}
H_{\rm nQRM}=\tilde{\nu}\adaga+\frac{\tilde{\omega}}{2}\sigma_z+g_n [e^{i\phi_n}\sigma^+(a^n+(\adag)^n)+{\rm H.c}]
\end{equation}
with $\phi_n=n\pi/2$ and $g_n=\eta^n\Omega/(2 \ n!)$. The validity of Eq.~(\ref{eq:nQRM}) is ensured when $\Omega\ll\nu$,   $|\tilde{\omega}+n \tilde{\nu}|\ll n\nu$ together with $|\eta|\sqrt{\left<(a+\adag)^2 \right>}\ll 1$ to safely perform a rotating wave approximation (RWA) in the joint  Hilbert space involving spin and bosonic degrees of freedom. In this respect, an expression of the leading order error committed by our scheme can be found in Sec. I of Supplementary Information~\cite{sup}.  The simulated nQRM can be brought into strong or ultrastrong coupling regimes as the parameters $\tilde{\omega}$ and $\tilde{\nu}$ can be tuned to frequencies comparable to $g_n$.

In this manner, having access to $H_{\rm gQRM}$ that includes only a linear spin-boson interaction, enables the exploration of a nQRM with nonlinear spin-boson coupling ($n>1$), whose physics is fundamentally different. For example, the most exotic hallmarks of the two-photon QRM (2QRM), are that the spectrum becomes a continuum for $g_2=\tilde{\nu}/2$ regardless of $\tilde{\omega}$, and for $g_2>\tilde{\nu}/2$ the Hamiltonian is not longer lower bounded~\cite{Chen:12,Peng:13,Felicetti:15,Duan:16}.  The gQRM lacks these features, and it is therefore not obvious that the physics of  $H_{\rm 2QRM}$ can be accessed  from $H_{\rm gQRM}$.  Moreover, the $H_{\rm gQRM}$ allows to simulate more exotic scenarios like combined nQRM and mQRM (see Sec. II in Supplementary Information~\cite{sup}).

Based on the previous transformations one can find the following expression among operators that establishes a relation between the gQRM and nQRM dynamics, which is the central result of this article (see Methods for a more detailed derivation):

\begin{align}
\label{eq:UgQRM}
\mathcal{U}_{\rm gQRM}\approx \Gamma^{\dagger}(t) \mathcal{U}_{\rm nQRM} T^{\dagger}(i\eta/2).
\end{align}

Here, $\mathcal{U}_{\rm gQRM}$ and $\mathcal{U}_{\rm nQRM}$ are the propagators of the gQRM and nQRM respectively,  $\Gamma(t)=\mathcal{U}_{s,0}^{\dagger}T^{\dagger}(i\eta/2)\mathcal{U}_{T,0}$ with $\mathcal{U}_{s,0}=e^{-itH_{s,0}}$, and the approximate character of Eq.~(\ref{eq:UgQRM}) is only a consequence of the RWA performed to achieve $H_{\rm nQRM}$ from $H_s$. Hence, an initial state $\left|\psi_{\rm nQRM}(0) \right>$ after an evolution time $t$ under $H_{\rm nQRM}$ can be approximated as
$\left|\psi_{\rm nQRM}(t) \right>\approx \Gamma(t)\left|\psi_{\rm gQRM}(t) \right>$ 
with the initial state $\left|\psi_{\rm gQRM}(0) \right>=T(i\eta/2)\left|\psi_{\rm nQRM}(0) \right>$. 

Remarkably, while the dynamics under the gQRM occurs in a typical time $1/(\eta\nu)$, see Eq.~(\ref{eq:gQRM}), the simulated nQRM (Eq.~(\ref{eq:nQRM})) involves parameters that are much smaller than $\nu$  since they satisfy the previously commented conditions $\Omega\ll\nu$, $|\tilde{\omega}+n\tilde{\nu}|\ll n\nu$, and $g_n=\eta^n\Omega/(2 \ n!)$.  As a consequence, a long evolution time of gQRM is required to effectively reconstruct the dynamics of nQRM.

Finally, our theory is completed with a mapping for the observables. As it can be derived from Eq.~(\ref{eq:UgQRM}) (see Methods), the expectation value of an observable $O_{\rm nQRM}$, i.e. an observable of the nQRM, corresponds to evaluate $O_{\rm gQRM}=\Gamma^{\dagger}(t)O_{\rm nQRM}\Gamma(t)$ in the gQRM. Because $\Gamma(t)$ involves bosonic displacement and spin rotations, $O_{\rm gQRM}$ may be in general intricate. Yet, for two relevant observables in nQRM, $\sigma_z$ and $\adaga$, the mapping leads to simple operators, namely,  $\sigma_z$ transforms into $-\sigma_x$ and $\adaga$ into $\adaga-\eta/2 p\sigma_x +\eta^2/4$ (see Methods). Interestingly, it still possible to obtain good approximations for other observables by truncating bosonic operators. Indeed, $e^{-\eta^2/2}\left[\sigma_{z,y}\cos((\tilde{\omega}+\delta_1)t)\mp\sigma_{y,z}\sin((\tilde{\omega}+\delta_1)t)\right]$ turns to be a good approximation of $\sigma_{x,y}$ in the gQRM frame (see Sec. III in Supplementary Information~\cite{sup}) which allows to recover the full qubit dynamics of nQRM. 

\begin{figure}
  \centering
\includegraphics[width=0.86\linewidth,angle=-90]{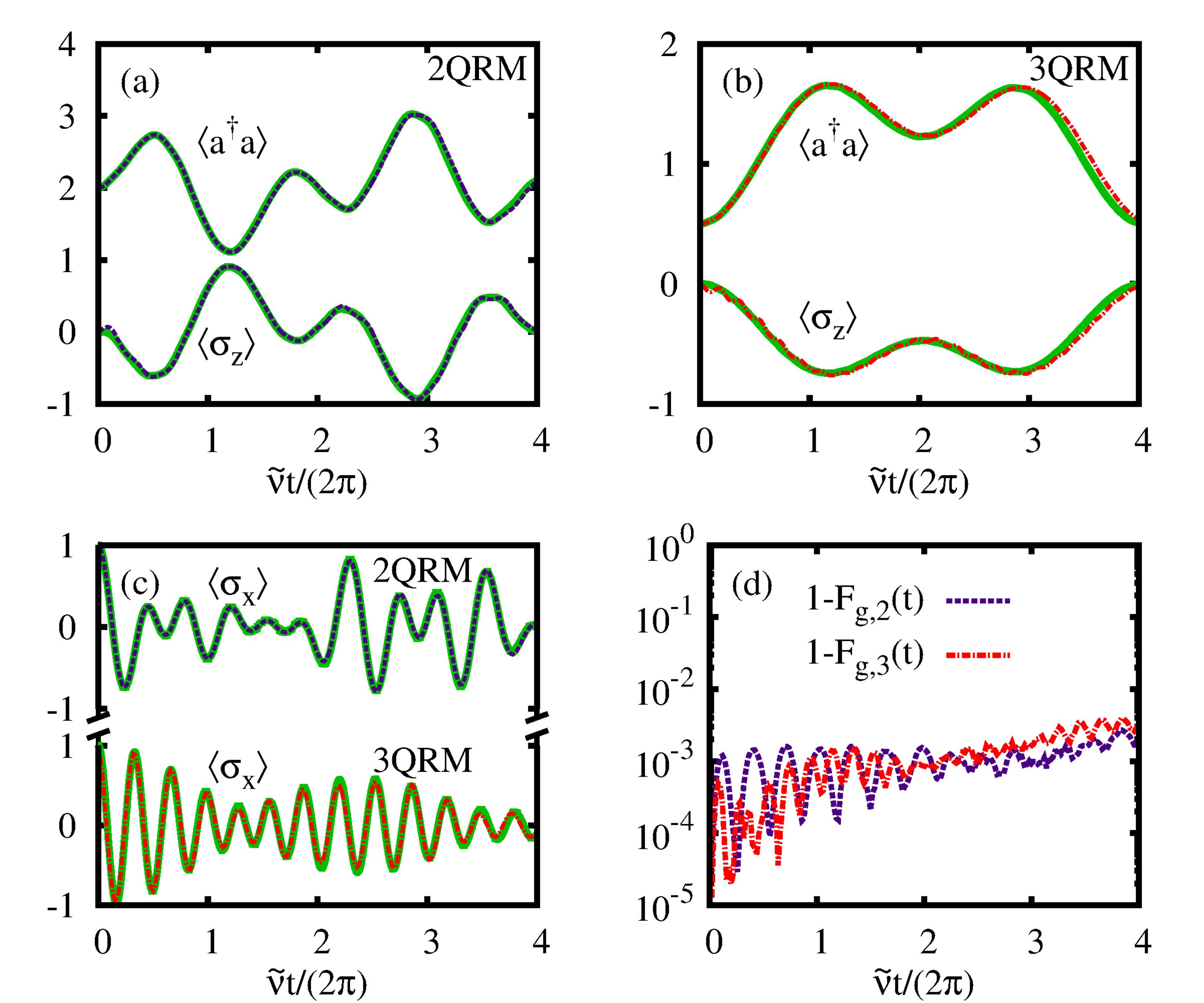}
\caption{\small{Simulated $nQRM$ dynamics using $gQRM$. Comparison between the dynamics of $H_{\rm nQRM}$ and the simulated one using $H_{\rm gQRM}$, for $n=2$ (a) and $n=3$ (b). The panels (a) and (b) show $\left<\sigma_z\right>$ and $\left< \adaga\right>$ of the nQRM (solid green lines) and their counterpart in the gQRM frame, that is, $-\left<\sigma_x\right>$ and $\left<\adaga\right>-\eta/2\left<p\sigma_x \right>+\eta^2/4$, respectively, depicted with dashed dark blue (2QRM) and dotted-dashed red lines (3QRM). With the same style, in (c) we show the ideal $\left<\sigma_x\right>$ for 2QRM (top) and 3QRM (bottom) and its approximated counterpart using gQRM, obtained through bosonic truncation (see main text).  We take as initial state $\left|\psi_{\rm 2QRM}(0)\right>=\left|2\right>\left|\uparrow\right>_x$ and the parameters $g_2/\tilde{\nu}=0.125$ and $\tilde{\omega}=2\tilde{\nu}$.  For the 3QRM, $\left|\psi_{\rm 3QRM}(0)\right>=(\ket{0}+\ket{1})/\sqrt{2}\left|\uparrow\right>_x$, $g_3/\tilde{\nu}=0.05$ and $\tilde{\omega}=3\tilde{\nu}$.  For $H_{\rm gQRM}$, $\omega/\nu=10^8$, $\Omega/\nu=0.1$ and $\tilde{\nu}/\nu=5\times 10^{-4}$. In (d) we show the infidelity between the states for the two considered models, $1-F_{\rm g,2}(t)$ (dashed dark blue line)  and $1-F_{\rm g,3}(t)$ (dotted-dashed red line).} }
\label{fig1}
\end{figure}

\subsection*{Approximate equivalence among {\rm gQRM} and {\rm 2} and {\rm 3QRM}}

To numerically confirm our approximate equivalence, in Fig.~\ref{fig1} we show the results of the simulated dynamics of a 2QRM and a 3QRM using a gQRM for a certain set of parameters and initial states $\left|\psi(0)_{\rm 2QRM}\right>=\left|2\right>\left|\uparrow\right>_x$ and $\left|\psi(0)_{\rm 3QRM}\right>=(\ket{0}+\ket{1})/\sqrt{2}\left|\uparrow\right>_x$, with $\left|\uparrow(\downarrow) \right>_x=(\left|e\right>\pm\left|g\right>)/\sqrt{2}$. In addition, in Fig.~\ref{fig1}(c) we show that the targeted $\sigma_x$ of a nQRM is retrieved by means of the previously mentioned bosonic truncation of $\sigma_x$ in the gQRM frame, i.e., $e^{-\eta^2/2}\left[\sigma_{z}\cos((\tilde{\omega}+\delta_1)t)-\sigma_{y}\sin((\tilde{\omega}+\delta_1)t)\right]$. Furthermore, in order to quantify the agreement among these models and the validity of the previous theory, we compute the fidelity between the ideal quantum state of the nQRM and the approximated state evolved in the gQRM and properly transformed with $\Gamma(t)$, that is, $F_{\rm g,n}(t)=\left<\psi_{\rm gQRM}(t)\right|\Gamma^{\dagger}(t)\left|\psi_{\rm nQRM}(t)\right>$. The computed fidelities of the considered cases are well above $0.99$, showing the good agreement among these two models. Note that although $H_{\rm 3QRM}$ could present truncation problems for $g_3\neq 0$ (see~\cite{Lo:98}), these do not affect the dynamics for the particular case plotted in Fig.~\ref{fig1}. Indeed, for the chosen parameters and initial state, the dynamics  during the considered evolution takes place in a constrained  region of the Hilbert space and thus it does not show Fock space truncation problems (see Sec. IV in Supplementary Information~\cite{sup} for further details). It is however worth stressing that this is not the general case, because the 3QRM is not bounded from below. Therefore, the number of excitations can grow very fast and, as a consequence, the simulation of the 3QRM relying on the approximate equivalence will break down since $\eta\sqrt{\left<\left(a+\adag\right)^2\right>}\ll 1$ is not longer satisfied. It is important to mention that our approximate equivalence is, in addition, not restricted to small times. The latter assertion is corroborated in Fig.~\ref{fig1} where the propagators for the 2QRM and 3QRM for the final time $t_f=2\pi \frac{4}{\tilde \nu}$ (values of $\tilde{\nu}$ in the caption) are  $\exp{\left\{ -i\pi [\frac{\tilde{\nu}}{g_2} a^\dag a + \frac{\tilde{\omega}}{2g_2}\sigma_z - \sigma_x (a^2 + (a^\dag)^2)]\right\}}$ and $\exp{\left\{ -i2\pi/5 [\frac{\tilde{\nu}}{g_3} a^\dag a + \frac{\tilde{\omega}}{2g_3}\sigma_z + \sigma_y (a^3 + (a^\dag)^3)]\right\}}$ respectively. Note that in both previous cases the coupling terms are multiplied by a phase  $\pi$ and $2\pi/5 $ respectively. We furthermore stress that these phases ($\pi$ and $2\pi/5$) can be increased without deteriorating the achieved fidelities by simply choosing a larger value for $\nu$. As previously commented, this is indeed possible since the approximate character of our method appears when we equal $H_s$ to $H_{\rm nQRM}$, whose performance is enhanced for large values of $\nu$ (see Methods).

\subsection*{Application for microwave driven ions}
The proof-of-concept of our method can be illustrated in a microwave driven ions platform.  Note that the developed theory may be relevant in other systems as circuit QED~\cite{Devoret:13}.  A microwave-driven trapped ion in a magnetic field gradient is described by (for more details see \cite{Mintert:01,Timoney:11,Weidt:16,Piltz:16})

\begin{align}\label{microwave}
H_{\rm MW} =& \frac{\omega}{2}\sigma_z + \nu a^{\dag}a\nonumber\\&+\Delta (a+a^\dag)\sigma_z+\sum_j\Omega_j \sigma_{x} \cos{(\omega_jt + \varphi_j)},
\end{align}
where $\omega$ is the qubit energy splitting with a value that depends on the ion species. For example, for $^{171}$Yb$^+$, we have $\omega \approx 12.4$ GHz~\cite{Olmschenk:07} plus a factor $\gamma B_z$ with $\gamma \approx 1.4$ MHz/G that depends on the applied static magnetic field $B_z$. 
The coupling parameter $\Delta$ determines the rate of the spin-boson coupling, while the last term corresponds to the action of microwave radiation on the system~\cite{Arrazola:17}.  
In this setup the spin-boson coupling is restricted to be linear, and therefore our theory appears as an alternative to introduce higher-order boson couplings in the dynamics.  In order to take Eq.~(\ref{microwave}) into the form of Eq.~(\ref{eq:gQRM}), and subsequently (via the mapping $T$) into the general  expression in Eq.~(\ref{eq:Hs}), we define $\omega = \delta_1 + \tilde{\omega}$ and move to a rotating frame with respect to the term $\frac{\tilde{\omega}}{2}\sigma_z$. Considering two drivings such that $\varphi_{1,2} =\pi$, $\omega_1 = \tilde{\omega}$ and $\omega_2 = \tilde{\omega} - (\delta_2 - \delta_1)$ and after eliminating terms that rotate at frequencies on the order of GHz, we find 

\begin{align}\label{eq:mwh}
H^I_{\rm MW}=&\nu a^\dag a + \Delta (a+a^\dag)\sigma_z +\frac{\delta_1}{2}\sigma_z\nonumber\\&-\frac{\Omega}{2}\sigma_x - \frac{\Omega}{2}(\sigma^+ e^{i(\delta_2-\delta_1)t} + \rm{H.c.}),
\end{align}
which equals $H_{\rm gQRM}$ after a basis change, that is, $e^{-i\frac{\pi}{4} \sigma_y} e^{-i\frac{\pi}{2} a^\dag a}H_{\rm MW}^I e^{i\frac{\pi}{2} a^\dag a}e^{i\frac{\pi}{4} \sigma_y} = H_{\rm gQRM},$
where $H_{\rm gQRM}$ is given in Eq.~(\ref{eq:gQRM}) with $\eta=2\Delta/\nu$. Hence, it is possible to use a microwave-driven ion to simulate models with nonlinear spin-boson couplings (see Sec. V in Supplementary Information~\cite{sup} for more details concerning the implementation in this setup). 

\begin{figure}
\centering
\includegraphics[width=0.43\linewidth,angle=-90]{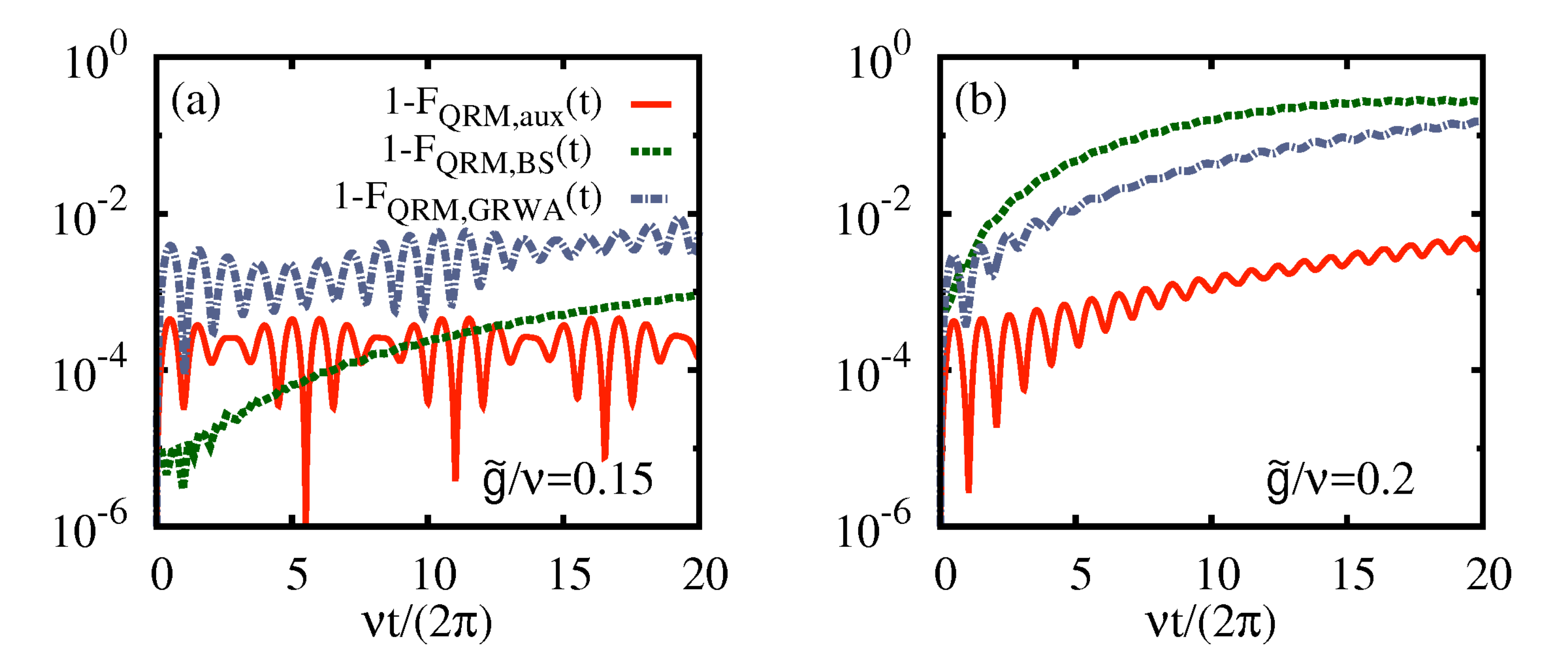}
\caption{\small{Approximate solutions to the QRM. Infidelity between the time-evolved state of QRM and its approximate solution evolved from $H_{\rm aux}$, the BS Hamiltonian $H_{\rm BS}$ and the GRWA approach $H_{\rm GRWA}$. These are denoted by $1-F_{\rm QRM,aux}$ (solid red), $1-F_{\rm QRM,BS}$ (dashed green) and $1-F_{\rm QRM,GRWA}$ (dashed-dotted blue), respectively. For (a)  $\Omega/\nu=0.1$, $\tilde{g}/\nu=\eta/2=0.15$ and $\left|\psi(0) \right>=\left|0\right>\ket{\uparrow}_x$, and in (b) $\Omega/\nu=0.04$, $\eta/2=0.2$ and $\left|\psi(0) \right>=\left|2\right>\left|g\right>$.}} 
\label{fig2}
\end{figure}

\subsection*{Approximate analytical solution for the {\rm QRM}.}

Finding a solution to the QRM has been subject of a long-standing debate, which still attracts considerable attention~\cite{Braak:11,Chen:12,Zhong:13,Batchelor:15}. Based on our theory, we obtain a simple expression for the  time-evolution propagator and expectation values of the QRM. The general expression given in Eq.~(\ref{eq:gQRM}) adopts the form of a standard QRM with a unique driving and $\delta_{1}=0$,
\begin{align}
\label{eq:QRM}
H_{\rm QRM}=H_{\rm gQRM}(\delta_1=0)=\nu\adaga-\frac{\eta \nu}{2}p\sigma_x+\frac{\Omega}{2}\sigma_z, 
\end{align}
which, applying our method, approximately corresponds to $H_{\rm aux}=\Omega/2\sigma_x[1-\eta^2(\adaga+1/2)]$. Indeed, from $H_s=H_{s,0}+H_{s,1}$ with $H_{s,0}=\nu\adaga+\omega\sigma_z/2$ (setting $\tilde{\omega}=\tilde{\nu}=0$), we obtain now $\mathcal{U}^{\dagger}_{s,0}H_{s,1}\mathcal{U}_{s,0}\approx H_{\rm aux}$ instead of $H_{\rm nQRM}$, and
where fast oscillating terms have been neglected performing a RWA, requiring again $|\eta| \sqrt{\left<(a+\adag)^2\right>}\ll 1$, and only considering resonant terms up to $\eta^2$ (see Sec. IV in Supplementary Information~\cite{sup}). As a consequence,  the following analysis does not apply to the deep-strong coupling regime~\cite{Casanova:10}, found here when $\eta \geq 2$.
Hence, the propagator for the QRM is approximated as

\begin{align}
\label{eq:UQRM}
&\mathcal{U}_{\rm QRM}\approx \mathcal{U}_{T,0}^{\dagger}T(i\eta/2)\mathcal{U}_{s,0} \mathcal{U}_{\rm aux}T^{\dagger}(i\eta/2),
\end{align}
which is expected to hold even in the ultrastrong coupling regime of the QRM, although restricted to the condition $\Omega\ll \nu$.
Because $H_{\rm aux}$ has a simple form, the time evolution can be analytically solved, with an initial state $\ket{\psi_{\rm aux}(0)}=T^{\dagger}(i\eta/2)\ket{\psi_{\rm QRM}(0)}$. Indeed, $H_{\rm aux}=\sum_{n,\pm} E_n^{\pm}\ket{\varphi_n^{\pm}}\bra{\varphi_n^{\pm}}$ with $\ket{\varphi_n^{\pm}}=\ket{n}\ket{\uparrow (\downarrow)}_x$ and $E_n^{\pm}=\pm \Omega/2(1-\eta^2(n+1/2))$. Now, employing the map between the two models, Eq.~(\ref{eq:UQRM}), we obtain the relation between observables. For example,  $\adaga$ in the QRM translates to  $\adaga+\eta^2/4+\eta/2(x\sigma_z\sin\nu t-p\sigma_z\cos\nu t)$ in $H_{\rm aux}$ (see Sec. VII in Supplementary Information~\cite{sup}).
In addition, we show that our method improves the typical Bloch-Siegert (BS) approximation~\cite{Beaudoin:11,Rossatto:16} and the generalised RWA (GRWA) of the QRM~\cite{Feranchuk:96,Irish:07,Gan:10} in a particular parameter regime. The former, i.e. the BS, is found as $e^{-S}H_{\rm QRM}e^{S}\approx H_{\rm BS}$, with $H_{\rm BS}=(\nu+\tilde{g}\Lambda\sigma_z)\adaga+(\Omega+\tilde{g}\Lambda)/2 \sigma_z-\tilde{g}(i\adag \sigma^--ia\sigma^+)$, where the anti-Hermitian operator is given by $S=i \Lambda(\adag \sigma_++a\sigma_-)-\xi\sigma_z(a^2-(a^{\dagger})^2)$, with parameters $\Lambda=\tilde{g}/(\nu+\Omega)$, $\xi=\tilde{g}\Lambda/(2\nu)$ and $\tilde{g}=\eta\nu/2$ (see~\cite{Beaudoin:11,Rossatto:16}). The GRWA of the QRM is attained in a similar manner, but with $S=\tilde{g}/\nu\chi \sigma_z(a^{\dagger}-a)$ such that $e^{-S}H_{\rm QRM}e^{S}\approx H_{\rm GRWA}$ where $H_{\rm GRWA}$ has a Jaynes-Cummings form with modified parameters, see~\cite{Feranchuk:96,Irish:07,Gan:10} and Sec. VIII in the Supplementary Information~\cite{sup} for further details.  In Fig.~\ref{fig2} we compute the overlap between time-evolved states for these three approaches (our approximate solution, the BS approximation, and the GRWA) and the QRM.  The approximate solution reproduces correctly the time evolution of the QRM as the coupling enters in the non-perturbative ultrastrong regime, $\tilde{g}/\nu=0.2$ (see~\cite{Rossatto:16}) with a fidelity $F_{\rm QRM,aux}>0.99$, while approximations $H_{\rm BS}$ and $H_{\rm GRWA}$ fail as their fidelities drop significantly. For smaller couplings these approaches lead to similar high fidelities (see Fig.~\ref{fig2}(a)).


\section*{Discussion}

We have presented a connection, i.e. an approximate equivalence, among a family of Hamiltonians, including the QRM and its higher order counterparts (nQRM) comprising a nonlinear interaction term that involves the simultaneous exchange of $n$ bosonic excitations with the spin-qubit, such as the two-photon QRM.  In particular, the standard QRM including spin driving terms, i.e. the gQRM, allows us to retrieve the nQRM dynamics with very high fidelities.  This theoretical framework shows that  nQRMs can be accessed even in the absence of the required nonlinear spin-boson exchange terms, as illustrated with a microwave-driven trapped ion. Therefore, we find that this fundamental model, the gQRM,  approximately contains the dynamics of all other nth order models.
  Moreover, we have derived an approximate solution to the dynamics of the QRM even in the ultrastrong coupling regime which surpasses in accuracy previous approximate solutions.   In this manner,  we have defined a general theoretical frame for the study and understanding of this family of fundamental Hamiltonians and their associated dynamics, which may open new avenues in quantum computing and simulation.

\section*{Methods}

\subsection*{Transformation between $H_s$, $H_{\rm gQRM}$ and $H_{\rm nQRM}$}

The Hamiltonian $H_s$, given in Eq.~(\ref{eq:Hs}), after the unitary transformation $H_T=T(i\eta/2)H_{s}T^{\dagger}(i\eta/2)$, adopts the following form
\begin{align}
H_T=&\nu\adaga-\frac{\omega}{2}\sigma_x+\frac{i\eta\nu}{2}(a-\adag)\sigma_x+\frac{\nu\eta^2}{4}\nonumber\\&+\frac{\Omega}{2}\sum_j \left[ \cos\alpha_j\ \sigma_z+\sin\alpha_j\ \sigma_y\right],\label{eq:HTmethods}
\end{align}
which becomes $H_{\rm gQRM}$ in the rotating frame with respect to $H_{T,0}=-(\omega+\delta_1)\sigma_x/2$, namely, $H_{\rm gQRM}\equiv \mathcal{U}_{T,0}^{\dagger}(t)(H_T-H_{T,0})\mathcal{U}_{T,0}(t)$ as given in Eq.~(\ref{eq:gQRM}). For simplicity, we constrain ourselves to the case in which $\Omega_j\equiv \Omega \ \forall j$, although the procedure can be easily extended to a more general scenario. 
On the other hand, $H_s$ leads to the desired nQRM when moving to an interaction picture with respect to $H_{s,0}=(\nu-\tilde{\nu})\adaga+(\omega-\tilde{\omega})/2\sigma_z$ with $H_s=H_{s,0}+H_{s,1}$
Then, the interacting part of $H_s$ can be written as 
\begin{align}
H_{s,1}^I&\equiv \mathcal{U}_{s,0}^{\dagger}(t,t_0)(H_{s}-H_{s,0})\mathcal{U}_{s,0}^{\dagger}(t,t_0)\nonumber\\ &=\tilde{\nu}\adaga+\frac{\tilde{\omega}}{2}\sigma_z\nonumber\\&\quad+\sum_j\frac{\Omega}{2}\left\{ \sigma^+e^{i(\omega-\tilde{\omega}) t'}e^{i\eta(a(t')+\adag(t'))}e^{-i\alpha_j}+{\rm H.c.}\right\},
\end{align}
with $a(t)=a e^{-i(\nu-\tilde{\nu})t}$, $\adag(t)=\adag e^{i(\nu-\tilde{\nu}) t}$ and $\mathcal{U}_{s,0}(t,t_0)$ the time-evolution operator associated to $H_{s,0}$ such that $t'=t-t_0$. Then, expanding the exponential, considering that $\Omega\ll \nu$ and $|\eta| \sqrt{\left<(a+\adag)^2 \right>}\ll 1$, and that $\delta_{1,2}=\mp n\nu-\tilde{\omega}\pm n\tilde{\nu}$ with $|\tilde{\omega}+n\tilde{\nu}|\ll n\nu$, one can perform a rotating wave approximation just keeping those terms resonant with $\sigma^+a^{n}$ and $\sigma^-a^{n}$.  In general, 
\begin{align}
H_{s,1}^I\approx & H_{\rm nQRM}=\tilde{\nu}\adaga+\frac{\tilde{\omega}}{2}\sigma_z\nonumber\\&\quad+g_n[e^{i\phi_n}\sigma^+  + e^{-i\phi_n}\sigma^-] \times [a^n  + (a^\dag)^{n}],\label{eq:Hnmethods}
\end{align}
with $g_n=\eta^n\Omega/(2\ n!)$ and $\phi_n=n\pi/2$. Hence, it is possible to achieve a $H_{\rm nQRM}$ from $H_s$. Note however that the corresponding attained coupling $g_n$ becomes smaller for increasing $n$, as it is proportional to $\eta^n/n!$.  In particular, for $n=2$, $H_{s,1}^I$ can be approximated as
\begin{align}
H_{s,1}^I&\approx H_{\rm 2QRM}= \tilde{\nu}\adaga+\frac{\tilde{\omega}}{2}\sigma_z-\frac{\eta^2\Omega}{4}\sigma_x \left( a^2 + (a^\dag)^2\right).
\end{align}

Note that, while the Hamiltonians $H_s$ and  $H_{\rm gQRM}$ are related through a unitary transformation, the achievement of a n-photon QRM, $H_{\rm nQRM}$, from $H_s$ requires of certain relations between parameters, such as $\Omega\ll \nu$, $|\tilde{\omega}+n\tilde{\nu}|\ll n\nu$ and  $|\eta| \sqrt{\left<(a+\adag)^2 \right>}\ll 1$ to safely perform the rotating wave approximation. In addition, it is worth stressing that the  equivalence to a good approximation  is not restricted to $H_{\rm nQRM}$ and $H_{\rm gQRM}$. For example,  a $H_{\rm gQRM}$ can lead into a more complex Hamiltonian, such as one comprising both nQRM and mQRM interaction terms (see Supplementary Information~\cite{sup}).

\subsection*{Transformations of observables and states}

Here we show the derivation of the Eq.~(\ref{eq:UgQRM}) which is a central result of this article. Having established the transformations that connect $H_{\rm gQRM}$ with $H_{s}$, and $H_{\rm nQRM}$ with $H_{s}$ we can relate them in terms of the time-evolution operators, 
\begin{align}
\label{eqs:Us1}
\mathcal{U}_{T}&=T(i\eta/2)\mathcal{U}_s T^{\dagger}(i\eta/2)\\\label{eqs:Us2}
\mathcal{U}_{T}&=\mathcal{U}_{T,0}\mathcal{U}^I_{T,1}=\mathcal{U}_{T,0}\mathcal{U}_{\rm gQRM}\\\label{eqs:Us3}
\mathcal{U}_{s}&=\mathcal{U}_{s,0}\mathcal{U}_{s,1}^I\approx \mathcal{U}_{s,0}\mathcal{U}_{\rm nQRM}
\end{align}
where $\mathcal{U}^I_{x,1}$ denotes the time-evolution propagator of $H_{x,1}$ in an interaction picture with respect to $H_{x,0}$ such that $H_{x}=H_{s,0}+H_{s,1}$. Note that we have dropped  the explicit time dependence for the sake of readability (see previous Eqs.~(\ref{eq:HTmethods})-~(\ref{eq:Hnmethods}) for the specific transformations). Then, combining the Eqs.~(\ref{eqs:Us1}), (\ref{eqs:Us2}) and (\ref{eqs:Us3}), we arrive to 
\begin{align}
\mathcal{U}_{\rm gQRM}\approx \mathcal{U}_{T,0}^{\dagger}T(i\eta/2) \mathcal{U}_{s,0}\mathcal{U}_{\rm nQRM}T^{\dagger}(i\eta/2)
\end{align}
which is the Eq.~(\ref{eq:UgQRM}), $\mathcal{U}_{\rm gQRM}=\Gamma^{\dagger}(t)\mathcal{U}_{\rm nQRM}T^{\dagger}(i\eta/2)$ with $\Gamma(t)=\mathcal{U}_{s,0}^{\dagger}T^{\dagger}(i\eta/2)\mathcal{U}_{T,0}$. Then, 
\begin{align}
\left|\psi_{\rm nQRM}(t)\right>&=\mathcal{U}_{\rm nQRM}\left|\psi_{\rm nQRM}(0)\right>\nonumber\\&\approx \Gamma(t) \mathcal{U}_{\rm gQRM} T(i\eta/2)\left|\psi_{\rm nQRM}(0)\right>\nonumber\\&=\Gamma(t)\left|\psi_{\rm gQRM}(t)\right>\label{eqs:ngT}
\end{align}
with the relation between initial states $\left|\psi_{\rm gQRM}(0)\right>=T(i\eta/2)\left|\psi_{\rm nQRM}(0)\right>$. Finally, from Eq.~(\ref{eqs:ngT}) it is straightforward to obtain the observable that must be measured in the gQRM frame in order to retrieve $O_{\rm nQRM}$ of the nQRM, i.e., $O_{\rm gQRM}=\Gamma^{\dagger}(t)O_{\rm nQRM}\Gamma(t)$. Explicitly, $\Gamma(t)$ reads
\begin{align}
\Gamma(t)=e^{-it(\tilde{\omega}-\omega)/2\sigma_z}e^{-it(\tilde{\nu}-\nu)\adaga}T^{\dagger}(i\eta/2)e^{-it(-(\omega+\delta_1)/2\sigma_x)}\nonumber
\end{align}
and thus, for $O_{\rm nQRM}=\sigma_z$ and $\adaga$ the transformation leads to
\begin{align}
(\sigma_z)_{\rm gQRM}&=-\sigma_x\\
(\adaga)_{\rm gQRM}&=\adaga-\frac{\eta}{2}p\sigma_x+\frac{\eta^2}{4},
\end{align}
while for other observables, like $\sigma_x$ and $\sigma_y$, a more intricate expression is attained,
\begin{widetext}
\begin{align}
\label{eqs:sxgQRM}
(\sigma_x)_{\rm gQRM}&=\left\{\cos((\omega+\delta_1)t)\sigma_z-\sin((\omega+\delta_1)t)\sigma_y \right\}{\rm Re}\left[ \mathcal{D}(i\eta)e^{i(\omega-\tilde{\omega})t}\right]\nonumber\\
&+\left\{\sin((\omega+\delta_1)t)\sigma_z+\cos((\omega+\delta_1)t)\sigma_y \right\}{\rm Im}\left[ \mathcal{D}(i\eta)e^{i(\omega-\tilde{\omega})t}\right]\\
\label{eqs:sygQRM}
(\sigma_y)_{\rm gQRM}&=\left\{\sin((\omega+\delta_1)t)\sigma_z+\cos((\omega+\delta_1)t)\sigma_y \right\}{\rm Re}\left[ \mathcal{D}(i\eta)e^{i(\omega-\tilde{\omega})t}\right]\nonumber\\
&-\left\{\cos((\omega+\delta_1)t)\sigma_z-\sin((\omega+\delta_1)t)\sigma_y \right\}{\rm Im}\left[ \mathcal{D}(i\eta)e^{i(\omega-\tilde{\omega})t}\right]
\end{align}
\end{widetext}
as it involves  qubit and bosonic operators due to the presence of the displacement operator $\mathcal{D}(\beta)$. However, because the condition $|\eta|\sqrt{\left<(a+\adag)^2\right>}\ll 1$ is required to guarantee a good realisation of $H_{\rm nQRM}$ and so that of Eq.~(\ref{eq:UgQRM}), the previous expression can be well approximated by truncating $\mathcal{D}(\beta)$. Indeed,  in our case $\mathcal{D}(i\eta)$ can be approximated up to third order as
\begin{align}
\label{eqs:Dieta}
\mathcal{D}(i\eta)=&e^{-\eta^2/2}\left[I+i\eta(a+\adag)\right.\nonumber\\&\left.-\frac{\eta^2}{2}\left(2\adaga+(\adag)^2+a^2\right) +\mathcal{O}\left(\eta^3a^3 \right) \right].
\end{align} 
In general, we can approximate the observable  $(\sigma_j)_{\rm gQRM}$ by truncating at order $M$, that is,
\begin{align}
  \label{eqs:sxM}
(\sigma_j)_{\rm gQRM}\approx(\sigma_j)_{\rm gQRM}^M= \sum_{n=0}^M (\sigma_j)_{\rm gQRM}^{(n)},
\end{align}
where the terms $(\sigma_j)_{\rm gQRM}^{(n)}$ for $j=x,y$ and can be calculated from Eqs.~(\ref{eqs:sxgQRM}),~(\ref{eqs:sygQRM}) and~(\ref{eqs:Dieta}). In particular, for $\sigma_{x,y}$ and for $n=0$,
\begin{align}
\label{eqs:sx0}
&(\sigma_x)_{\rm gQRM}^{(0)}=\nonumber\\&\quad e^{-\eta^2/2}\left[\sigma_z \cos((\tilde{\omega}+\delta_1)t)-\sigma_y\sin((\tilde{\omega}+\delta_1)t)  \right],
\end{align}
\begin{align}
\label{eqs:sy0}
&(\sigma_y)_{\rm gQRM}^{(0)}=\nonumber\\&\quad e^{-\eta^2/2}\left[\sigma_z \sin((\tilde{\omega}+\delta_1)t)+\sigma_y\cos((\tilde{\omega}+\delta_1)t)  \right].
\end{align}
Note that measuring $(\sigma_{x,y})_{\rm gQRM}^{(M)}$ would require measurements of observables in the gQRM of the form $\sigma_{y,z}(a^M+(\adag)^M)$ as well as $\sigma_{y,z}(a^{\dagger})^na^m$ with $n+m=M$ and $n\geq m$  (see Supplementary Information~\cite{sup}). Remarkably, for the considered cases here, the zeroth order approximation already reproduces reasonably well the expectation value of $\sigma_{x,y}$ of a nQRM. Therefore, having access to qubit observables in gQRM, $\sigma_{x,y,z}$, allows to reconstruct the full qubit dynamics of a nQRM. Note that Eqs.~(\ref{eqs:sx0}) and~(\ref{eqs:sy0}) correspond to the expressions given in Results, which for $\sigma_x$ is plotted in Fig.~\ref{fig1}(c) for the simulation of a 2QRM and 3QRM.


\section*{Acknowledgements}
This work was supported by the ERC Synergy grant BioQ, the EU STREP project EQUAM. The authors acknowledge support by the state of Baden-W\"urttemberg through bwHPC and the German Research Foundation (DFG) through grant no INST 40/467-1 FUGG. J. C. acknowledges Universit\"at Ulm for a Forschungsbonus and support by the Juan de la Cierva grant IJCI-2016-29681. H. M.-C. thanks the Alexander von Humboldt Foundation for support. R. P. acknowledges DfE-SFI Investigator Programme (grant 15/IA/2864). J. C. and R. P. have contributed equally to this work.

%

\section*{Author contribution}
J.C. and R.P. conceived the idea  and develop the theory with inputs from H. M.-C. and M. B. P. All authors contributed to the writing of the manuscript.


\newpage
\widetext

\begin{center}
  \textbf{\large{Supplemental Information}}
\end{center}
\setcounter{equation}{0} \setcounter{figure}{0} \setcounter{table}{0}
\setcounter{page}{1} \makeatletter \global\long\def\theequation{S\arabic{equation}}
 \global\long\def\thefigure{S\arabic{figure}}
 \global\long\def\bibnumfmt#1{[S#1]}

\subsection*{I. Simulation of $H_{\rm nQRM}$:  analytical expression for the leading order error of the method and numerical analysis}

The approximate character of the equivalence in the main text appears when Eq.~(1) is approximated to Eq.~(3). Here, terms that contain nth order of bosonic operators are quasi-resonant, i.e. they are detuned by a small quantity $\propto n\tilde{\nu}$, while the rest of the terms are detuned by  $\propto n \nu$. Note that $|\nu| \gg |\tilde{\nu}|$. Among these highly detuned terms, the ones with the highest influence are those in which the $\eta$ parameter does not appear. More specifically these can be collected in the following Hamiltonian $Q$

\begin{equation}\label{eq:QSI}
Q = \frac{\Omega}{2}\sum_j (\sigma^+ e^{-i\delta_j t} +  \sigma^- e^{+i\delta_j t}),
\end{equation} 
where each $\delta_j$ is $\propto n {\nu}$ with $n$ the order of the target nQRM. One can calculate that the propagator associated to $Q$ up to first order in $(1/\delta_j)$ reads 

\begin{equation}
\mathcal{U}^Q_{[t:t_0]} \approx e^{i (t-t_0)\sum_j \frac{\Omega^2}{4 \delta_j} \sigma_z} + O((1/\delta_j)^2). 
\end{equation}
In this manner, the introduced error is always small if the coefficients $\frac{(t-t_0) \Omega^2}{4\delta_j}$  are small. The latter gets certified if  $\frac{\Omega}{\nu} \rightarrow 0$, note again that  $\delta \propto n \nu$. Follows a numerical analysis of the simulation of the 2QRM relying on the reported method.

\begin{figure}
\centering
\includegraphics[width=0.5\linewidth,angle=-90]{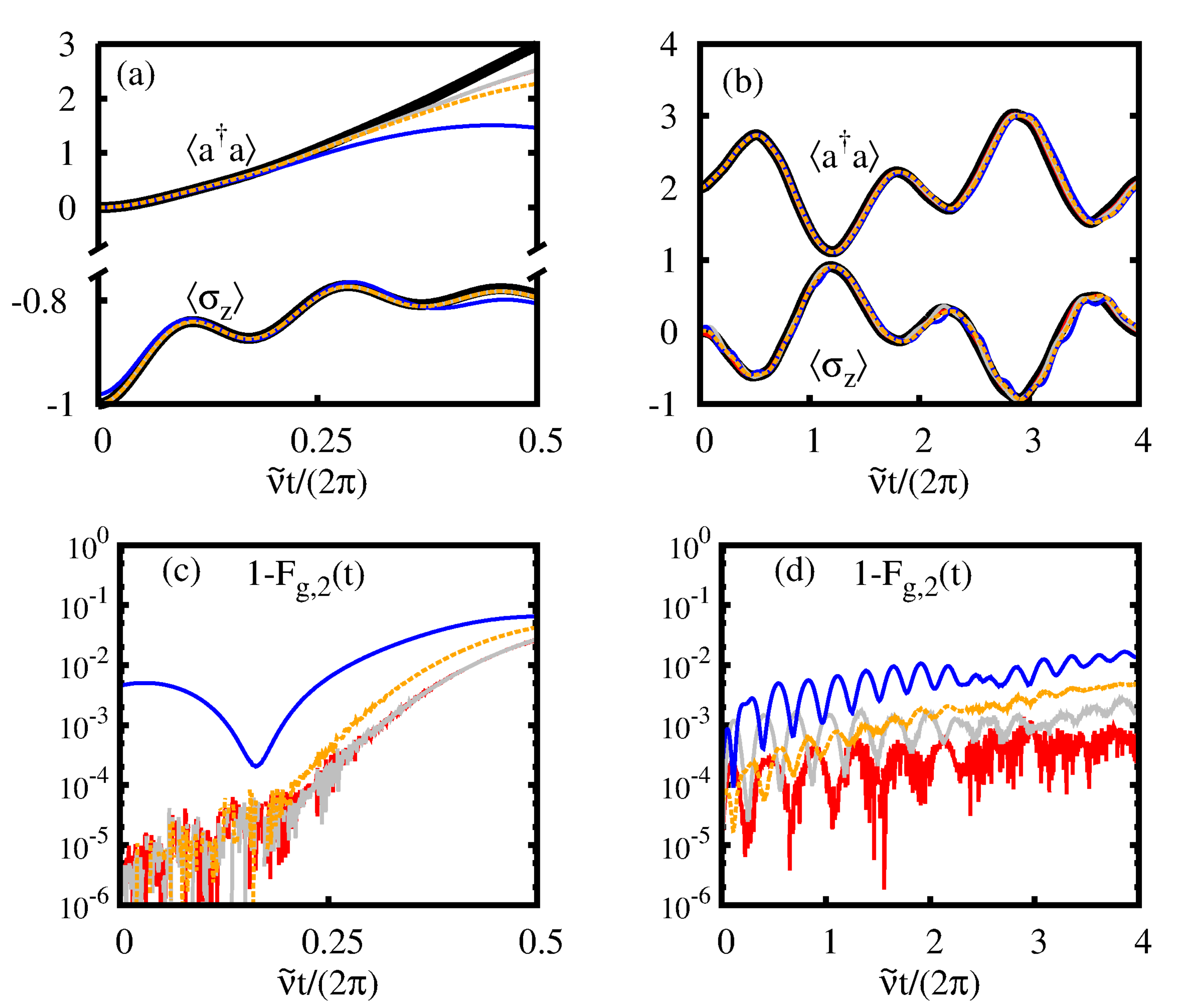}
\caption{\small{Simulated 2QRM dynamics for $g_2/\tilde{\nu}=1/2$ (spectral collapse) in (a) with $\ket{\psi_{\rm 2QRM}(0)}=\ket{0}\ket{g}$, and $g_2/\tilde{\nu}=0.125$ in (b) with $\ket{\psi_{\rm 2QRM}(0)}=\ket{2}\ket{\uparrow}_x$ for different parameters in gQRM, and their corresponding infidelities, $1-F_{\rm g,2}$, with respect to the 2QRM in (c) and (d), respectively.  The solid black lines correspond to the real 2QRM dynamics, while in (a) and (c), $\tilde{\nu}/\nu=1\times10^{-4}$, $\Omega/\nu=0.1$ (red and grey line), $\tilde{\nu}/\nu=1\times10^{-3}$, $\Omega/\nu=0.2$ (blue line) and $\tilde{\nu}/\nu=1\times10^{-4}$, $\Omega/\nu=0.05$ (orange line). The difference among the red and grey lines resides in the truncation of the Hilbert space, for the latter $N_{max}=200$ and the half for the former. In (b) and (c),  $\tilde{\nu}/\nu=2.5\times 10^{-4}$ and $\Omega/\nu=0.05$ (red line), $\tilde{\nu}/\nu=5\times 10^{-4}$ and $\Omega/\nu=0.1$ (grey line), $\tilde{\nu}/\nu=1\times 10^{-3}$ and $\Omega/\nu=0.2$ (blue line) and $\tilde{\nu}/\nu=1\times 10^{-4}$ and $\Omega/\nu=0.05$ (orange line). For all the cases, $\omega/\nu=10^8$.  } }
\label{fig1SM}
\end{figure}

As explained and detailed in the main text, the simulation of a nonlinear nQRM can be achieved from gQRM by properly choosing system's parameters. However, because the frequencies $\tilde{\omega}$ and $\tilde{\nu}$ can be tuned, different combinations of $\Omega$, $\nu$, $\tilde{\nu}$ and $\eta$ in $H_{\rm gQRM}$ can lead to the same simulated nQRM (see Eqs.~(3) and (11) of the main text). Therefore, it is important to recognise the main contribution that deteriorates the established approximate equivalence as it may be overcome by correctly tuning these free parameters.
Moreover, besides the chosen parameters, the simulation of the dynamics of the 2QRM when a large number of bosonic excitations is involved is expected to breakdown as $|\eta|\sqrt{\left<(a+\adag)^2\right>}\ll 1$ is not longer satisfied. This is indeed the case for the 2QRM right at the spectral collapse as $\left<\adaga\right>$ blows up, however, the onset of the dynamics can be still reproduced. In Fig.~\ref{fig1SM} we show a comparison of the simulated 2QRM with different parameters and for $g_2/\tilde{\nu}=1/2$ (at the spectral collapse) in (a) and (c), and $g_2/\tilde{\nu}=0.125$ in (b) and (d), as shown in Fig.~2 of the main text. Note that decreasing $\tilde{\nu}$ the real evolution time becomes longer, although and at the same time, it leads to smaller values of $\eta$ (and/or $\Omega/\nu$). For example, in Fig.~\ref{fig1SM}(d) we observe that the fidelity when $\Omega/\nu=0.05$ and $\tilde{\nu}/\nu=2.5\times 10^{-4}$ slightly improves that of $\Omega/\nu=0.1$ and $\tilde{\nu}/\nu=5\times 10^{-4}$, which together with the worst shown case ($\Omega/\nu=0.2$), share the same value of $\eta$, namely, $\eta=0.05$. This indicates that the main spurious contribution stems from the zeroth order in $\eta$, i.e.,  $\Omega/2 \sigma^{+}e^{\pm 2(\nu-\tilde{\nu}) t}$ (as given in Eq.~(\ref{eq:QSI})), as higher orders become smaller, $\propto \eta^n\Omega/(2 \ n!)$,  while rotating at approximately equal frequency, $\approx \nu$. In addition, we explicitly show that the presented results are not affected by the Fock-space truncation; in particular, for the spectral collapse, the results do not change when doubling the number of Fock states, from $N_{max}=100$ to $200$ (see Fig.~\ref{fig1SM}(a) and (c)).

\subsection*{II. Combined nth and mth order QRM models from $H_{\rm gQRM}$}
As stated in the main text, the approximate equivalence is not restricted to $H_{\rm nQRM}$ and $H_{\rm gQRM}$. Indeed, more complex Hamiltonians than nQRM can be attained by a suitable $H_{\rm gQRM}$. Here we show how a Hamiltonian that comprises interaction terms of that of a nQRM and a mQRM can be accessed from $H_{\rm gQRM}$. This Hamiltonian is denoted here by $H_{n,m}$ and reads

\begin{align}
H_{n,m}=\tilde{\nu}\adaga+\frac{\tilde{\omega}}{2}\sigma_z+\left[ g_ne^{i\phi_n}\sigma^+(a^n+(a^{\dagger})^n)+g_m e^{i\phi_m}\sigma^+(a^m+(a^{\dagger})^m)+{\rm H.c.}\right].
\end{align}
We first show how to achieve $H_{n,m}$ from $H_s$ (see Eq.~(1)), following the same procedure as shown for $H_{\rm nQRM}$. For that, four interaction terms are now needed in Eq.~(1), with $\alpha_j=\omega t+\delta_j t$ and $\delta_j=\mp n\nu -\tilde{\omega}\pm n\tilde{\nu}$ for $j=1, 2$ and $\delta_j=\mp m\nu -\tilde{\omega}\pm m\tilde{\nu}$ for $j=3, 4$. Then, assuming  that $\Omega\ll \nu$, $|\tilde{\omega}+n\tilde{\nu}|\ll n\nu$, $|\tilde{\omega}+m\tilde{\nu}|\ll m\nu$ as well as $|\eta|\sqrt{\left<a+\adag \right>}\ll 1$,  $H_{s,1}^I$ approximately corresponds to $H_{n,m}$, 

\begin{align}
H_{s,1}^I&\approx \tilde{\nu}\adaga+\frac{\tilde{\omega}}{2}\sigma_z+g_n \left[e^{i\phi_n}\sigma^++e^{-i\phi_n}\sigma^-\right]\times \left[a^n +(\adag)^n \right]+g_m \left[e^{i\phi_m}\sigma^++e^{-i\phi_m}\sigma^- \right]\times \left[a^m +(\adag)^m \right],
\end{align}
where we have dropped out non-resonant terms performing a rotating wave approximation (RWA), i. e., terms rotating at frequencies $\sim\nu$ have been neglected. Note that $H_{s,1}^I$ denotes $H_{s,1}$ in the interaction picture with respect to $H_{s,0}$ with $H_{s,0}=(\nu-\tilde{\nu})\adaga+(\omega-\tilde{\omega})\sigma_z/2$ and $H_s=H_{s,0}+H_{s,1}$, as explained in the main text and in the previous section. In addition, the attained phases are $\phi_k=k\pi/2$, while the couplings $g_k=\eta^k\Omega/(2 \ k!)$. It is worth mentioning that in this case one would gain tunability in the couplings by  considering different frequencies $\Omega$ for each $j$; for example, one could achieve similar couplings $g_n\sim g_m$ with $n\neq m$. 

On the other hand, following the procedure explained in the main text, we can bring $H_s$ into $H_{\rm gQRM}$, Eq.~(2), by applying a unitary transformation. For the particular case considered here, i.e., four drivings with $\alpha_j=\omega t+\delta_j t$ and $\delta_j=\mp n\nu -\tilde{\omega}\pm n\tilde{\nu}$ for $j=1, 2$ and $\delta_j=\mp m\nu -\tilde{\omega}\pm m\tilde{\nu}$ for $j=3, 4$, $H_{\rm  gQRM}$ adopts the following form

\begin{align}
H_{\rm gQRM}&=\nu \adaga+\frac{n(\tilde{\nu}-\nu)-\tilde{\omega}}{2}\sigma_x-\frac{\eta \nu}{2}p\sigma_x +\frac{\Omega}{2}\sigma_z+\frac{\Omega}{2}\left[\cos(2n(\nu-\tilde{\nu})t)+2\cos(n(\nu-\tilde{\nu})t)\cos(m(\nu-\tilde{\nu})t) \right]\sigma_z\nonumber\\&\qquad\qquad+\Omega\sin(n(\nu-\tilde{\nu})t)\left(\cos(n(\nu-\tilde{\nu})t)+\cos(m(\nu-\tilde{\nu})t) \right)\sigma_y.
\end{align}
Finally, the time-evolution propagators of both models are related as given in Eq.~(4), that is,

\begin{align}
\mathcal{U}_{\rm gQRM}\approx \Gamma^{\dagger}(t)\mathcal{U}_{n,m} T^{\dagger}(i\eta/2)
\end{align}
where $\Gamma(t)$ is defined as in the main text, $\Gamma(t)=\mathcal{U}^{\dagger}_{s,0}T^{\dagger}(i\eta/2)\mathcal{U}_{T,0}$. Recall that the unitary transformation $T(\beta)$ reads

\begin{align}\label{eq:TSI}
  T(\beta)=\frac{1}{\sqrt{2}}\left(
  \begin{matrix} \mathcal{D}^{\dagger}(\beta) & \mathcal{D}(\beta) \\ -\mathcal{D}^{\dagger}(\beta) & \mathcal{D}(\beta) \end{matrix} \right),
\end{align} where $\mathcal{D}(\beta)=e^{\beta a^{\dagger}-\beta^*a}$ is the usual displacement operator. 
Therefore, the map between initial states and observables is identical as explained in the main text for the approximate equivalence between a nQRM and a gQRM. Therefore, the observables of $H_{n,m}$ transform in the same manner to the frame of gQRM.

\begin{figure}
\centering
\includegraphics[width=0.25\linewidth,angle=-90]{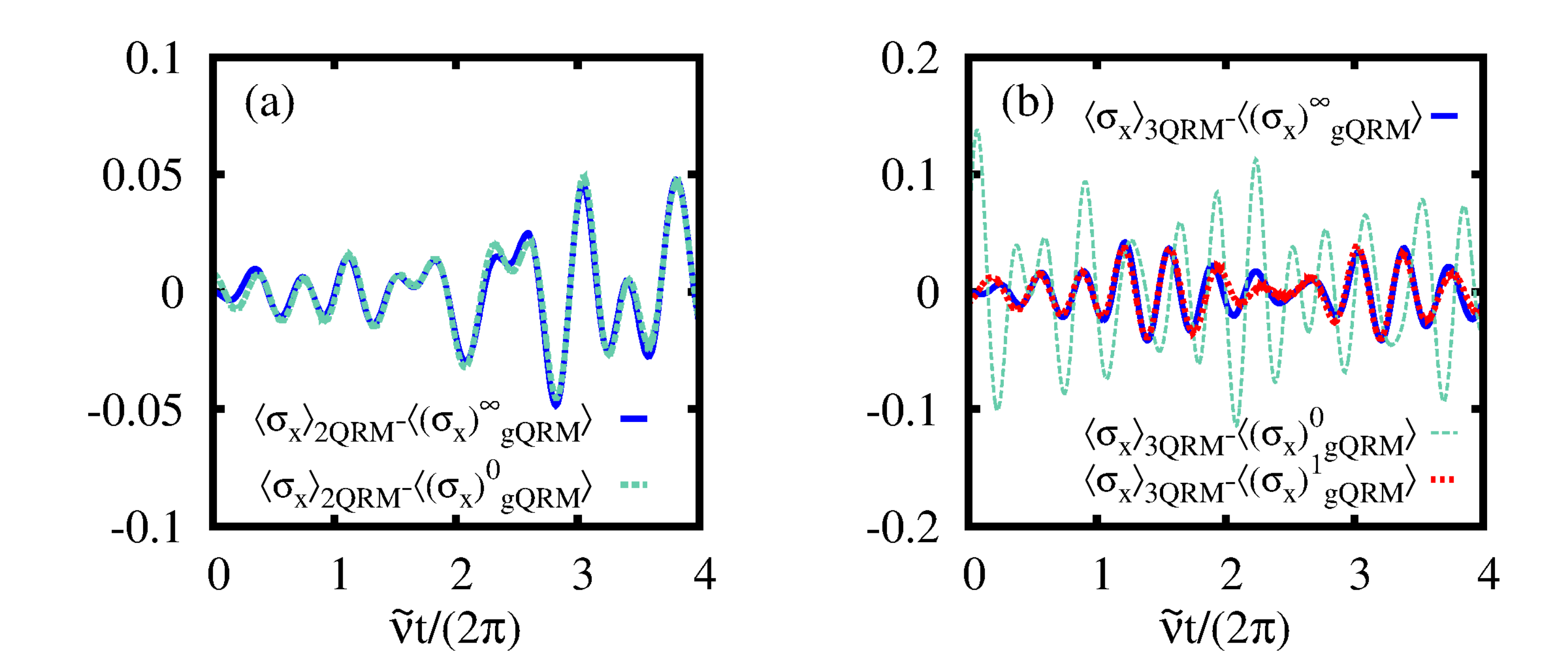}
\caption{\small{Difference between the ideal expectation value of $\sigma_x$ of a 2QRM (a) and a 3QRM (b), and its reconstruction using a gQRM truncating bosonic operators at zeroth order (dashed blue-green line), i.e., $(\sigma_x)_{\rm gQRM}^0$, first order (dashed red line) $(\sigma_x)_{\rm gQRM}^1$ (only for 3QRM) and without truncation (solid blue line), $(\sigma_x)_{\rm gQRM}^\infty$. The parameters and initial state are the same as in Fig.~2 of the main text. For the 2QRM, the initial state $\left|\psi(0)_{\rm 2QRM}\right>=\left|2\right>\left|\uparrow\right>_x$ and $g_2/\tilde{\nu}=0.125$ and $\tilde{\omega}=2\tilde{\nu}$, while for the 3QRM we consider $\left|\psi(0)_{\rm 3QRM}\right>=\left( \left|0 \right>+\left|1\right>\right)\left|\uparrow\right>_x$, $g_3/\tilde{\nu}=0.05$, and $\tilde{\omega}=3\tilde{\nu}$.  For $H_{\rm gQRM}$, $\omega/\nu=10^8$, $\Omega/\nu=0.1$, and $\tilde{\nu}/\nu=5\times 10^{-4}$. Note that $\left<\sigma_x\right>_{\rm nQRM}-\left<(\sigma_x)_{\rm gQRM}^\infty\right>$ is not zero due to the approximate character of the equivalence.}}
\label{fig2SM}
\end{figure}

\subsection*{III. Truncation of $\mathcal{D}(i\eta)$ for spin $\sigma_{x,y}$ observables}

As we have shown in the main text, an observable $O_{\rm nQRM}$ in the nQRM frame corresponds to $O_{\rm gQRM}=\Gamma^{\dagger}(t)O_{\rm nQRM}\Gamma(t)$ in the gQRM frame with (see Methods in the main text)

\begin{align}
\Gamma(t)=e^{-it(\tilde{\omega}-\omega)/2\sigma_z}e^{-it(\tilde{\nu}-\nu)\adaga}T^{\dagger}(i\eta/2)e^{-it(-(\omega+\delta_1)/2\sigma_x)}\nonumber
\end{align}
where $T(\beta)$ is given in Eq.~(\ref{eq:TSI}). Then, as stated in the main text, while for certain observables, the transformation leads to simple expressions, such as $O_{\rm nQRM}=\sigma_z$ or $\adaga$ (see Eqs.~(18) and (19) of main text), a more intricate form follows for $\sigma_x$ and $\sigma_y$. Indeed, they transform according to

\begin{align}
\label{eqs:sxgQRM}
(\sigma_x)_{\rm gQRM}&=\left\{\cos((\omega+\delta_1)t)\sigma_z-\sin((\omega+\delta_1)t)\sigma_y \right\} {\rm Re}\left[ \mathcal{D}(i\eta)e^{i(\omega-\tilde{\omega})t}\right]\nonumber\\
&\qquad\qquad\qquad+\left\{\sin((\omega+\delta_1)t)\sigma_z+\cos((\omega+\delta_1)t)\sigma_y \right\}{\rm Im}\left[ \mathcal{D}(i\eta)e^{i(\omega-\tilde{\omega})t}\right]\\
\label{eqs:sygQRM}
(\sigma_y)_{\rm gQRM}&=\left\{\sin((\omega+\delta_1)t)\sigma_z+\cos((\omega+\delta_1)t)\sigma_y \right\} {\rm Re}\left[ \mathcal{D}(i\eta)e^{i(\omega-\tilde{\omega})t}\right]\nonumber\\
&\qquad\qquad\qquad-\left\{\cos((\omega+\delta_1)t)\sigma_z-\sin((\omega+\delta_1)t)\sigma_y \right\}{\rm Im}\left[ \mathcal{D}(i\eta)e^{i(\omega-\tilde{\omega})t}\right]
\end{align}
Note that the previous expressions involve  mixed qubit and bosonic operators due to the presence of the displacement operator $\mathcal{D}(\beta)$. However, we can still truncate the expansion of $\mathcal{D}(\beta)$ since the condition $|\eta|\sqrt{\left<(a+\adag)^2\right>}\ll 1$ is required to guarantee a good approximate equivalence between nQRM and gQRM. Hence, we can expand $\mathcal{D}(\beta)$ in a sum of terms, whose $n$th term is proportional to $\eta^n/n!$ and contains $n$-order bosonic operators, namely, $(\adag)^pa^q$ such that $p+q=n$.  As given in the main text, we consider an expansion of $\mathcal{D}(i\eta)$ up to third order, 

\begin{align}
\label{eqs:Dieta}
\mathcal{D}&(i\eta)\approx e^{-\eta^2/2}\left[I+i\eta(a+\adag)-\frac{\eta^2}{2}\left(2\adaga+(\adag)^2+a^2\right) \right]
\end{align} 

In general, we can approximate the observable  $(\sigma_j)_{\rm gQRM}$ by truncating at order $M$, that is,

\begin{align}
\label{eqs:sxM}
(\sigma_j)_{\rm gQRM}\approx(\sigma_j)_{\rm gQRM}^M= \sum_{n=0}^M (\sigma_j)_{\rm gQRM}^{(n)}
\end{align}
where the terms $(\sigma_j)_{\rm gQRM}^{(n)}$ for $j=x,y$ and can be calculated from Eqs.~(\ref{eqs:sxgQRM}),~(\ref{eqs:sygQRM}) and~(\ref{eqs:Dieta}). In particular, for $\sigma_x$ and for $n=0$, $1$ and $2$ we obtain

\begin{align}
\label{eqs:sx0}
(\sigma_x)_{\rm gQRM}^{(0)}&=e^{-\eta^2/2}\left[\sigma_z \cos((\tilde{\omega}+\delta_1)t)-\sigma_y\sin((\tilde{\omega}+\delta_1)t)  \right]\\
(\sigma_x)_{\rm gQRM}^{(1)}&=\eta e^{-\eta^2/2}\left[\sigma_z\sin((\tilde{\omega}+\delta_1)t)+\sigma_y\cos((\tilde{\omega}+\delta_1)t) \right](a+\adag)\\
(\sigma_x)_{\rm gQRM}^{(2)}&=\frac{\eta^2}{2}e^{-\eta^2/2}\left[\sigma_y\sin((\tilde{\omega}+\delta_1)t)-\sigma_z\cos((\tilde{\omega}+\delta_1)t) \right](2\adaga+a^2+(\adag)^2)
\end{align}
while for $\sigma_y$ the following expressions are attained

\begin{align}
\label{eqs:sy0}
(\sigma_y)_{\rm gQRM}^{(0)}&=e^{-\eta^2/2}\left[\sigma_z \sin((\tilde{\omega}+\delta_1)t)+\sigma_y\cos((\tilde{\omega}+\delta_1)t)  \right]\\
(\sigma_y)_{\rm gQRM}^{(1)}&=\eta e^{-\eta^2/2}\left[-\sigma_z \cos((\tilde{\omega}+\delta_1)t)+\sigma_y\sin((\tilde{\omega}+\delta_1)t)  \right](a+\adag)\\
(\sigma_y)_{\rm gQRM}^{(2)}&=-\frac{\eta^2}{2}e^{-\eta^2/2}\left[\sigma_z\sin((\tilde{\omega}+\delta_1)t)+\sigma_y\cos((\tilde{\omega}+\delta_1)t) \right](2\adaga+a^2+(\adag)^2).
\end{align}
It is worth noticing that $(\sigma_{x,y})_{\rm gQRM}^{(M)}$ would require measurements of observables in the gQRM of the form $\sigma_{y,z}(a^M+(\adag)^M)$ as well as $\sigma_{y,z}(a^{\dagger})^na^m$ with $n+m=M$ and $n\geq m$. As stated in the main text, the zeroth order already provides a good approximation of the corresponding observables $\sigma_{x,y}$ of the nQRM, as we have shown for $\sigma_x$ in Fig.~2(c) of the main text. Here we analyse the deviation between the ideal $\sigma_x$ of a 2QRM and a 3QRM and its corresponding approximation by truncating at different orders. For $\sigma_y$ similar results are obtained, although not explicitly shown here. In particular, in Fig.~\ref{fig2SM} we show these deviations for the same case considered in Fig.~2 of the main text. In Fig.~\ref{fig2SM}(a) we show the difference between the ideal $\sigma_x$ of a 2QRM and its approximation in the gQRM frame at zeroth order, $(\sigma_x)_{\rm gQRM}^0$, and without performing any truncation, i.e., $(\sigma_x)_{\rm gQRM}^\infty$ which corresponds to the Eq.~(\ref{eqs:sxgQRM}), or equivalently, to the Eq.~(\ref{eqs:sxM}) with $M=\infty$. In Fig.~\ref{fig2SM}(b) we show the same differences but now for a 3QRM and including the first order, $(\sigma_x)_{\rm gQRM}^1$. We note that, while for the considered parameters and initial state for the 2QRM the zeroth order approximation of $\sigma_x$ is already as good as including all of terms, Eq.~(\ref{eqs:sxgQRM}), first order correction does matter for the specific case considered here in a 3QRM. These results unveil that the small difference between $\sigma_x$ and its truncated approximation at an order $M$, $(\sigma_x)_{\rm gQRM}^M$, stems mainly from the approximate character of the equivalence (Eq.~(4) of main text) and not due to truncation, as we find the same deviation when all the orders are included $(\sigma_x)_{\rm gQRM}^\infty$, since $|\eta|\sqrt{\left<(a+\adag)^2 \right>}\ll 1$. Note that, for the situation considered in Fig.~\ref{fig2SM}, $\eta=0.05$ and $\eta=0.1442$ for the 2QRM and 3QRM, respectively.

\begin{figure}
\centering
\includegraphics[width=0.5\linewidth,angle=-90]{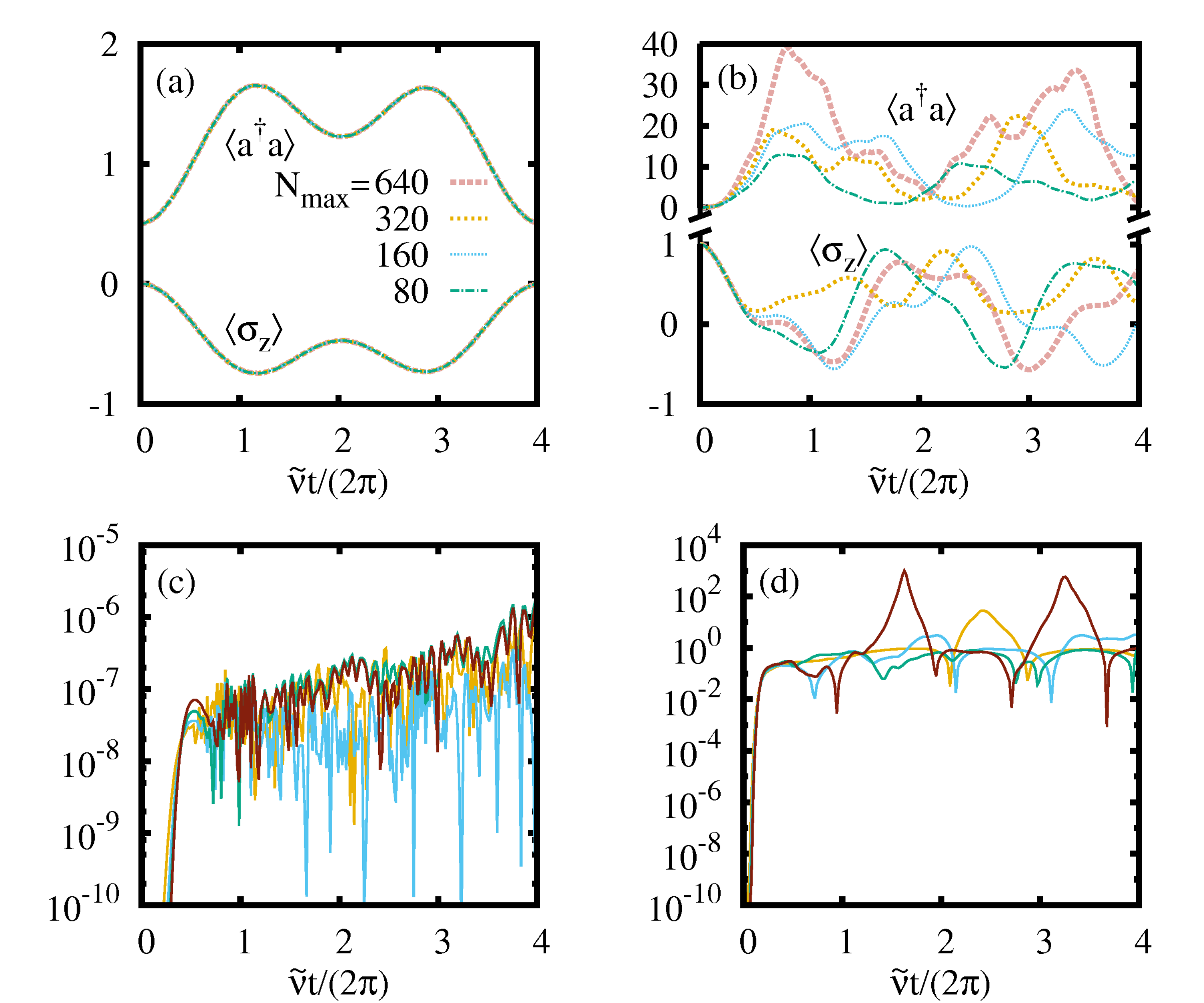}
\caption{\small{ Dynamics of a 3QRM at resonant condition $\tilde{\omega}=3\tilde{\nu}$ as a function of the maximum number of Fock states $N_{max}$. In (a), we plot the same condition considered in the main text, Fig. 2 (b), namely $g_3/\tilde{\nu}=0.05$ and $\ket{\psi_{\rm 3QRM}(0)}=\left(\ket{0}+\ket{1} \right)/\sqrt{2}\ket{\uparrow}_x$, which does not change increasing the truncation $N_{max}$. To the contrary, as illustrated in (b), for an initial state $\ket{\psi_{\rm 3QRM}(0)}=\ket{0}\ket{e}$ with  $g_3/\tilde{\nu}=0.15$ the dynamics blows up and thus the results are a numerical artefact because they strongly depend on $N_{max}$. In the lower panels, (c) and (d), we plot the relative difference $|\left<\adaga \right>_{N_{max}}-\left<\adaga \right>_{2N_{max}}|/\left<\adaga \right>_{2N_{max}}$ for the cases considered in (a) and (b), respectively. The curves depicted by orange, light blue, green and dark red lines correspond to $N_{max}=80$, $160$, $320$ and $640$. In (c) we observe that the relative difference does not  surpass $10^{-6}$, showing that the dynamics converge during the calculated time interval, while in (d), the difference largely increases as $N_{max}$ and no convergence is attained after a short evolution time. } }
\label{fig3SM}
\end{figure}

\subsection*{IV. Numerical results of the 3QRM}
  In the main text we have presented numerical results of the dynamics under a 3QRM, whose Hamiltonian can be written as (see Eq. (3) in the main text)

  \begin{align}
    H_{\rm 3QRM}=\tilde{\nu}\adaga+\frac{\tilde{\omega}}{2}\sigma_z+g_3\sigma_y\left(a^3+(\adag)^3\right).
  \end{align}
  As discussed in~\cite{Lo:98}, the previous Hamiltonian becomes unbounded from below for any $g_3\neq 0$. Yet, the time evolution of certain initial states does not diverge, i.e. their evolution remains in a small Hilbert space, during a time interval. Here we show that the dynamics of the 3QRM presented in the main text (Fig. 2)  is not a numerical artefact as a consequence of Hilbert space truncation, and thus, it might be accessed experimentally. In Fig.~\ref{fig3SM}, we show the expectation value of relevant observables when increasing the maximum number of considered Fock states $N_{max}$ for two different couplings $g_3$ and initial states at resonant condition, $\tilde{\omega}=3\tilde{\nu}$. We consider $N_{max}=80$, $160$, $320$, $640$ and $1280$. The analysis of the results strongly indicates that for $g_3/\tilde{\nu}=0.05$ and $\ket{\psi_{\rm 3QRM}(0)}=\left(\ket{0}+\ket{1} \right)/\sqrt{2}\ket{\uparrow}_x$, as done in the main text, the dynamics up to $t=4 \times 2\pi/\tilde{\nu}$ does converge. For comparison, we choose  $g_3/\tilde{\nu}=0.15$ and an initial state $\ket{\psi_{\rm 3QRM}(0)}=\ket{0}\ket{e}$ for which no convergence is attained after a very short evolution time. In order to illustrate the convergence, we plot the relative difference for the number of bosonic excitations as $N_{max}$ increases, namely, $|\left<\adaga \right>_{N_{max}}-\left<\adaga \right>_{2N_{max}}|/\left<\adaga \right>_{2N_{max}}$, where $\left<\adaga \right>_{N_{max}}$ denotes the expectation value of $\adaga$ with $N_{max}$ Fock states. For the case considered in the main text, this relative difference remains below $10^{-6}$. Despite this strong numerical evidence, a precise analysis regarding the convergence of the dynamics under $H_{\rm 3QRM}$ depending on evolution time, system's parameters and initial states remains to be disclosed.

\subsection*{V. Parameters for the implementation using microwave driven ions}
As commented in the main text, for case of a $^{171}$Yb$^+$ ion, the qubit energy splitting is $\omega \approx 12.4$ GHz~\cite{Olmschenk:07}, which is modified depending on the applied static magnetic field $B_z$ through a shift $\gamma B_z$ with $\gamma \approx 1.4$ MHz/G. 

The parameters for the results plotted in Fig.~2 of the main text are attainable with typical values in the $H_{\rm MW}$ setup. They can be realised with a trap frequency $\nu=2\pi\times 370$ kHz that, according to the ratio $\tilde{\nu}/\nu=5\times 10^{-4}$ leads to a maximum evolution time of $20$ ms, i.e. $\tilde{\nu}t/(2\pi)= 4$. In addition, for the case considered in Fig.~2(a), $\Delta$ has to be tuned to $\approx 9.25$ \ kHz which is achievable with a magnetic field gradient smaller than $150 \ \rm{\frac{T}{m}}$~\cite{Weidt:16}. Note however that an evolution time of $20$ ms is a rather long time to preserve the coherence of both qubit and bosonic mode  from the inevitable presence of environmental noise sources, even when techniques to cope with noise are applied (see~\cite{Piltz:16} where a coherence time of $\sim 10$ ms is measured). In this regard, we stress that although the results presented in Fig.~2 will be deteriorated by loss of coherence, a faithful simulation of nQRMs with $H_{\rm MW}$ is still feasible at shorter times. Furthermore, depending on the specific platform, the parameters may be optimised to avoid spurious decoherence processes or be combined with techniques to extend quantum coherence as dynamical decoupling~\cite{Puebla:16njp}. Finally, it is worth emphasising that we have considered a microwave driven ion just to illustrate a direct application of our theory, as a proof of concept.

\subsection*{VI. Relation between $H_{\rm QRM}$ and $H_{\rm aux}$} 
The Eq.~(2) of the main text reduces to a QRM by simply considering a driving with $\delta_1=0$, that is,

\begin{align}
H_{\rm QRM}=\nu\adaga-\frac{\eta\nu}{2}p\sigma_x+\frac{\Omega}{2}\sigma_z.
\end{align}
Making use of the derived approximate equivalence, the previous Hamiltonian can be approximately mapped into a simple Hamiltonian, $H_{\rm aux}=\frac{\Omega}{2}\left[1-\eta^2(\adaga+1/2)\right]$. This is accomplished by moving $H_{s}=H_{s,0}+H_{s,1}$ to an interaction picture with respect to $H_{s,0}=\nu\adaga+\omega\sigma/2$. Note that now $\tilde{\omega}=\tilde{\nu}=0$ and recall that $\alpha_1=(\omega+\delta_1)t=\omega t$. Therefore,

\begin{align}
H_{s,1}^I=\mathcal{U}_{s,0}^{\dagger}H_{s,1}\mathcal{U}_{s,0}=\frac{\Omega}{2}\left[\sigma^+ e^{i\eta\left( a(t)+\adag(t)\right)}+{\rm H.c.} \right],
\end{align}
where $a(t)=ae^{-i\nu t}$ and $\adag(t)=\adag e^{i\nu t}$. Requiring now $|\eta|\sqrt{\left<(a+\adag)^2\right>}\ll 1$ we expand the exponential, and assuming $\Omega\ll \nu$, we can safely perform a RWA neglecting off-resonant terms (which rotate at frequencies larger or equal than $\nu$) and keeping only the resonant terms up to $\eta^2$. The following higher-order resonant term appears with $\eta^4$. Hence, we obtain the relation given in the main text, 

\begin{align}
H_{s,1}^I\approx H_{\rm aux}=\frac{\Omega}{2}\left[1-\eta^2(\adaga+1/2)\right].
\end{align}
In a straightforward manner as done for nQRM and gQRM, we can obtain the relation between the propagators, $\mathcal{U}_{\rm QRM}$, $\mathcal{U}_{\rm aux}$ and $\mathcal{U}_{s,0}$, 

\begin{align}
&\mathcal{U}_{\rm QRM}\approx \mathcal{U}_{T,0}^{\dagger}T(i\eta/2)\mathcal{U}_{s,0} \mathcal{U}_{\rm aux}T^{\dagger}(i\eta/2),
\end{align}
Note that this relation also follows from Eq.~(4) considering now that gQRM reduces to QRM and $H_{\rm aux}$ replaces $H_{\rm nQRM}$, and with $H_{s,0}=\nu\adaga+\omega\sigma/2$ since $\tilde{\omega}=\tilde{\nu}=0$.

\subsection*{VII. Expectation values in the approximate QRM}
As indicated in the main text and from Eq.~(8),

\begin{align}
&\mathcal{U}_{\rm QRM}\approx \mathcal{U}_{T,0}^{\dagger}T(i\eta/2)\mathcal{U}_{s,0} \mathcal{U}_{\rm aux}T^{\dagger}(i\eta/2),
\end{align}
which relates the time evolution of a QRM with the simple evolution in the Hamiltonian $H_{\rm aux}$, one can obtain the map between observables and initial states. In particular, we have

\begin{align}
  \mathcal{U}_{T,0}=e^{-it(-\omega/2\sigma_x)}\qquad {\rm and}\qquad  \mathcal{U}_{s,0}=e^{-it(\omega/2\sigma_z+\nu\adaga)}
  \end{align}
where we have set already $\tilde{\omega}=\tilde{\nu}=0$ in $H_{s,0}$, i.e.,  $H_{s,0}=\nu\adaga+\omega\sigma_z/2$. In addition, $T(\beta)$ denotes the unitary transformation given in the main text, that is, $T(\beta)=1/\sqrt{2}\left[\mathcal{D}(\beta)\left(\ket{e}\bra{g}+\ket{g}\bra{g}\right)+\mathcal{D}^{\dagger}(\beta)\left(\ket{e}\bra{e}-\ket{g}\bra{e}\right) \right]$ with $\mathcal{D}(\beta)=e^{\beta \adag-\beta^{*}a}$ the displacement operator. Then, it follows that the initial state transforms $\ket{\psi_{\rm aux}(0)}=T^{\dagger}(i\eta/2)\ket{\psi_{\rm QRM}(0)}$ and the observables $O_{\rm aux}=\mathcal{U}_{s,0}^{\dagger}T^{\dagger}(i\eta/2)\mathcal{U}_{T,0} O_{\rm QRM} \mathcal{U}_{T,0}^{\dagger}T(i\eta/2)\mathcal{U}_{s,0}$, which leads to

\begin{align}
  (\adaga)_{\rm QRM}&\rightarrow \adaga+\frac{\eta^2}{4}+\frac{\eta}{2}\left(x\sigma_z\sin\nu t-p\sigma_z\cos\nu t \right)\\
  (x)_{\rm QRM}&\rightarrow x\cos\nu t+p\sin \nu t\\
  (p)_{\rm QRM}&\rightarrow p\cos\nu t-x\sin\nu t-\eta\sigma_z,
  \end{align}
where the r.h.s corresponds to observables in the $H_{\rm aux}$ frame. The expectation value of these observables can now be computed in a straightforward manner, as illustrated here for $\left<x\sigma_z\right>$,

\begin{align}
\left<x\sigma_z\right>&=\left< \psi_{\rm aux}(t)\right| x\sigma_z\left| \psi_{\rm aux}(t)\right>=\sum_{n,l}\sum_{m,k}\bra{\varphi_n^{l}} (C_n^{l})^{*} e^{itE_n^{l}}x\sigma_z e^{-itE_m^{k}}C_m^{k}\ket{\varphi_m^k}\nonumber\\
&=\sum_{n,m}(C_n^{\mp})^*C_m^{\pm} e^{it(E_n^\mp-E_m^\pm)}\times\left( \sqrt{m+1}\delta_{n,m+1}+\sqrt{m}\delta_{n,m-1}\right)
\end{align}
where $C_n^{\pm}=\left<\varphi_{n}^\pm\right|\left.\psi_{\rm aux}(0) \right>$ corresponds to the expansion of the initial state in the eigenstates of $H_{\rm aux}$, i.e., $\ket{\varphi_{n}^{\pm}}=\ket{n}\ket{\uparrow(\downarrow)}_x$ with eigenvalues $E_n^{\pm}=\pm \Omega/2(1-\eta^2(n+1/2))$.

\subsection*{VIII. Bloch-Siegert approximation and generalised RWA of the QRM}
As commented in the main text, we compare the developed approximate solution of the QRM by means of our approximate equivalence with two customary procedures, namely, the Bloch-Siegert approximation $H_{\rm BS}$~\cite{Beaudoin:11,Rossatto:16} and the generalised RWA $H_{\rm GRWA}$~\cite{Feranchuk:96,Irish:07,Gan:10}. In the following we briefly summarise the main outcomes of these approaches, while referring to the interested reader to the previous references for further details.
As stated in the main text, the Bloch-Siegert approximation consists in transforming $H_{\rm QRM}$ (Eq. (7) in main text with $\tilde{g}=\eta\nu/2$) according to $e^{-S}H_{\rm QRM}e^{S}$ with $S=i\Lambda(\sigma_+a^{\dagger}+\sigma_- a)-\xi\sigma_z(a^2-(a^\dagger)^2)$, $\Lambda=\tilde{g}/(\nu+\Omega)$ and $\xi=\tilde{g}\Lambda/\nu$. Then, the transformed Hamiltonian corresponds to $H_{\rm BS}$ up to $\Lambda^2$,

\begin{align}\label{eqsup:bs}
H_{\rm BS}=(\nu+\tilde{g}\Lambda\sigma_z)\adaga+\frac{\Omega+\tilde{g}\Lambda}{2}\sigma_z-\tilde{g}(i\sigma_- a^{\dagger}-i\sigma_+ a).
\end{align}
It is straightforward to obtain the relation between both models, which follows from

\begin{align}\ket{\psi_{\rm QRM}(t)}=\mathcal{U}_{\rm QRM}\ket{\psi_{\rm QRM}(0)}\approx e^S \mathcal{U}_{\rm BS}e^{-S}\ket{\psi_{\rm QRM}(0)}=e^S \ket{\psi_{\rm BS}(t)}, \end{align}
with initial state $\ket{\psi_{\rm BS}(0)}=e^{-S}\ket{\psi_{\rm QRM}(0)}$. The generalised RWA approach first transforms the Hamiltonian $\tilde{H}_{\rm QRM}$ and then neglects counter-rotating terms and multiple-boson transitions. For the sake of simplicity we consider $\tilde{H}_{\rm QRM}=\nu\adaga+\frac{\Omega}{2}\sigma_x+\tilde{g}(a+\adag)\sigma_z$. Note that the Hamiltonian given in Eq. (7) is retrieved upon a rotation of spin and boson degrees of freedom, as shown after Eq. (6). The transformed Hamiltonian, $e^{-S} \tilde{H}_{\rm QRM}e^{S}\approx H_{\rm GRWA}$ where now $S=\tilde{g}/\nu \xi \sigma_z(a-\adag)$ with $\xi=(1+\beta \Omega/\nu)^{-1}$ and $\beta=e^{-4\tilde{g}^2\xi^2/\nu^2}$. The Hamiltonian $H_{\rm GRWA}$ adopts the form of a Jaynes-Cummings model up to constant factors,

\begin{align}H_{\rm GRWA}=\nu'\adaga+\frac{\Omega'}{2}\sigma_x+g'(\sigma_+a+\sigma_-\adag),  \end{align}
where the parameters are $\nu'=\nu$, $\Omega'=\beta\Omega$ and $g'=2\xi\beta\Omega\tilde{g}/\nu$. As stated in~\cite{Gan:10},  this method slightly improves the similar approximation developed earlier in~\cite{Feranchuk:96,Irish:07}.  Note that the relation between $H_{\rm GRWA}$ and a QRM immediately follows from Eq.~(\ref{eqsup:bs}) but with a different anti-Hermitian operator $S$. As our developed approximation $H_{\rm aux}$ holds for $\Omega\ll \nu$ (see main text) , the parameter $\xi$ can be well approximated by $\xi\approx (1+\Omega/\nu)^{-1}$. The performance of these approaches is shown in Fig. 3 of the main text, relying on the state fidelity between $\ket{\psi_{\rm QRM}(t)}$ and its approximated counterpart using $H_{\rm aux}$, $H_{\rm BS}$ and $H_{\rm GRWA}$.

\end{document}